\documentclass[12pt,preprint]{aastex}

\shorttitle{Transverse oscillations of a multi-stranded loop}

\shortauthors{M. Luna et al.}

\begin{document}

\title{Transverse oscillations of a multi-stranded loop}

\author{M. Luna\altaffilmark{1,2}, J.
Terradas\altaffilmark{1}, R. Oliver\altaffilmark{1}, and J.L.
Ballester\altaffilmark{1}}

\altaffiltext{1}{Departament de F\'{\i}sica, Universitat de les Illes Balears,
07122 Palma de Mallorca, Spain. Email: jaume.terradas@uib.es,
ramon.oliver@uib.es and joseluis.ballester@uib.es}
\altaffiltext{2}{CRESST and Space Weather Laboratory NASA/GSFC, Greenbelt, MD
20771, USA. Email: manuel.lunabennasar@nasa.gov}

\begin{abstract}
We investigate the transverse oscillations of a line-tied multi-stranded coronal
loop composed of several parallel cylindrical strands. First, the collective
fast normal modes of the loop are found with the $T$-matrix theory. There is a
huge quantity of normal modes with very different frequencies and a complex
structure of the associated magnetic pressure perturbation and velocity field.
The modes can be classified as bottom, middle, and top according to their
frequencies and spatial structure. Second, the temporal evolution of the
velocity and magnetic pressure perturbation after an initial disturbance are
analyzed. We find complex motions of the strands. The frequency analysis reveals
that these motions are a combination of low and high frequency modes. The
complexity of the strand motions produces a strong modulation of the whole tube
movement. We conclude that the presumed internal fine structure of a loop
influences its transverse oscillations and so its transverse dynamics  cannot be
properly described by those of an equivalent monolithic loop.
\end{abstract}

\section{Introduction}

Coronal loops are magnetic structures belonging to active regions in the solar
atmosphere. Observations with telescopes onboard the Solar Heliospheric
Observatory (SOHO), the Transition Region and Coronal Explorer (TRACE), and more
recently the Solar Terrestrial Relations Observatory (STEREO) and the HINODE
satellites show that such structures are flux tubes filled with plasma hotter
and denser than the surrounding corona. They are arches rooted in the
photosphere that outline the magnetic field. Nowadays, it is debated whether
coronal loops have an internal fine structure below the spatial resolution of
the current telescopes. In the so-called multi-stranded loop model it is
suggested that each observed loop is composed of a bundle of several tens or
hundreds of different strands \citep[see,
e.g.,][]{litwinrosner1993,aschwanden2000,klimchuk2006}. The internal fine
structure of loops allows to explain some observational aspects of loops. For
example, the uniform emission measure along loops \citep{lenz1999} was explained
assuming a multithermal internal structure \citep{reale2000}; in addition,
\citet{schmelz2001} argued that the broad differential emission measure is a
clear evidence of the multithermal structure of loops.

Transverse coronal loop oscillations, reported first in \citet{aschwanden1999},
were interpreted in terms of the fundamental kink mode \citep{nakariakov1999} of
a cylindrical loop with a uniform internal structure in the so-called monolithic
model \citep{edwin&roberts1983}. However, in the multi-stranded model of a loop,
the transverse motion of each strand can be influenced by the displacements of
its neighbors. Then, the internal fine structure can affect the oscillation
period, damping rate, and in general the dynamics of the whole loop. Thus, the
transverse oscillations of a multi-stranded loop can be different from those of
the monolithic tube model. Recently, \citet{ofman2008} described the first
indirect evidence of transverse oscillations in a multi-stranded coronal loop.
The authors also considered the loop as a collection of independent flux tubes.

Seismology of coronal loops \citep{uchida1970,roberts1984} relates the
observed properties of loop oscillations with theoretical models, and derives
local parameters that are difficult to measure directly. This method was
first applied to an observation of transverse loop oscillations by
\citet{nakariakov2001}, and who obtained an estimation of the magnetic field strength. Similarly,
\citet{wang2007} reported an observation of slow waves in coronal
loops, and used a seismological approach to deduce the field
strength. Both works compared observations with the straight cylindrical
model of \citet{edwin&roberts1983}. However, \citet{demoortel2009}
showed that local parameter estimation strongly depends on the theoretical
model used to compare with the observed system. The authors found discrepancies
of up to $50\%$ between the estimated and actual magnetic field when the
oscillatory parameters of a curved three-dimensional loop are compared
with those of a straight cylindrical tube.

For these reasons, a theoretical study of the transverse oscillations of a
multi-stranded loop model is necessary. An increasing number of publications
have considered the dynamics of flux tube ensembles. In \citet{berton1987} an
analytical investigation of the oscillations of a system of magnetic slabs
periodically distributed was made. In \citet{kris93,kris94} a qualitative study
of the wave propagation in a system of two slabs was considered. 
\citet{diaz2005} studied the oscillations of a prominence multifibril system
modeled as up to five non-identical slabs. In a system of two identical fibrils,
phase or antiphase oscillations were found, although the antiphase motions
rapidly leak their energy into the coronal medium. \citet{luna2006} studied a
system of two identical coronal slabs and found that the antiphase oscillations
can also be trapped. In addition, these authors found that after an initial
perturbation the system oscillates with a combination of the two collective
normal modes and a complex dynamics is produced. This study was extended to
cylindrical geometry in \citet{luna2008}, who considered two identical flux
tubes. Four trapped normal modes were found with frequencies different from
those of the individual tube. The time-dependent problem was numerically solved
and again a very complex dynamics associated to the mutual interaction of the
tubes was found. These studies show that the flux tube transverse oscillations
are coupled in systems of two identical loops. The dependence of the transverse
oscillations on the relative tube parameters, i.e. radii and densities, was
first studied in a system of two loops by \citet{doorsselaere2008}, who computed
the normal modes analytically with the long-wavelength approximation. However,
in \citet{luna2009} the normal modes of two and three loops were found
analytically. These authors found that the transverse oscillations of a set of
flux tubes are coupled if the kink frequency of the individual tubes are
similar, whereas the oscillations are uncoupled if they have sufficiently
different individual kink frequencies. \citet{arregui2007}, studied the effects
on the dynamics of the possibly unresolved internal structure of a coronal loop
composed of two very close, parallel, identical coronal slabs in Cartesian
geometry with non-uniform density in the transverse direction. They found small
differences in the period and damping time with respect to those of a single
slab with the same density contrast or a single slab with the same total mass.
\citet{ofman2005} investigated numerically the oscillations and damping
time of a nonlinear and highly resistive MHD model of four cylindrical strands.
This work was extended to a system of four strands with twist in
\citet{ofman2009}. \citet{terradas2008} numerically investigated the temporal
evolution of a system of $10$ strands with transverse non-uniform layers and
with smooth density profiles. They found that the system oscillates with a
global mode and that resonant absorption still provides a rapid and effective
damping of the loop transverse displacement.

The purpose of this work is to study the influence of the internal fine
structure of a loop on its transverse dynamics. We first compute analytically
the normal modes of different strand systems. We determine the different kinds
of collective normal modes and compare them with those of an monolithic flux
tube. We also study the temporal evolution of a system of ten identical strands
after an initial perturbation by solving numerically the initial value problem.
The results obtained are compared with those of the normal mode analysis.

The paper is arranged as follows. In \S \ref{model} the multi-stranded loop
model is presented and the equivalent monolithic loop is defined. We
analytically find the normal modes of ten identical strands in \S
\ref{10id_strands}, ten non-identical strands in \S \ref{non_id_strands}, and
forty identical strands in \S \ref{more_complex}. In \S \ref{time_evolution} the
initial value problem is numerically solved and the relation between the
temporal evolution with the normal modes is discussed. Finally, in \S
\ref{dis_conc} the results of this investigation are summarized and the
conclusions are drawn.

\section{Theoretical model}\label{model}

\begin{figure}[!ht]
\centering
\includegraphics[width=9.cm]{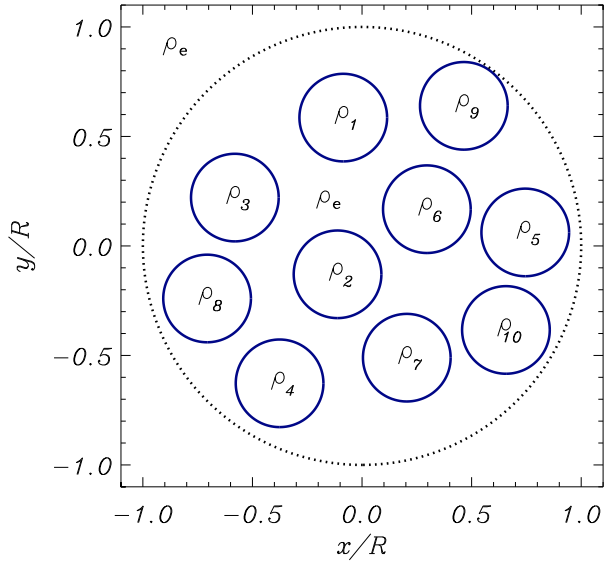} \caption{ Sketch of the cross
section of a multi-stranded loop model, which consists of a loop of radius $R$
(large dotted circle) filled with $N$ homogeneous strands of densities
$\rho_{j}$ and radii $a$ (solid smaller circles). The external medium to
the loop and the medium between strands consists of coronal material with
density
$\rho_\mathrm{e}$. It is important to note that the large dotted circle is not
real and
represents the external boundary of a hypothetical unresolved loop. }
\label{sketch_multistrand}
\end{figure}

In this work a coronal loop is assumed to have a composite structure of several
strands. Each coronal strand is modeled as a straight cylinder with uniform
density along the tube (gravity is neglected) with the loop feet tied in the
photosphere. Thus, the multi-stranded loop equilibrium configuration consists of
a bundle of $N$ cylindrical, parallel, homogeneous strands. The $z$-axis points
in the direction of the strand axes. All the strands have the same length, $L$,
and each individual strand, labeled $j$, is characterized by the position of its
center in the $xy$-plane, $\mathbf{r}_{j}=x_{j} \mathbf{e}_x+y_{j}
\mathbf{e}_y$, its radius, $a$, and its density, $\rho_{j}$. The position of
each strand is randomly generated within a hypothetical unresolved loop of
radius $R$ (see Figure \ref{sketch_multistrand}). The density of the coronal
environment is $\rho_\mathrm{e}$. The uniform magnetic equilibrium field is
$\mathbf{B}_{0}= B_{0} \mathbf{e}_z$ inside the strands and in the coronal
medium. We consider small-amplitude perturbations in this equilibrium and use
the linearized ideal magnetohydrodynamic equations in the zero-$\beta$ limit. A
harmonic time dependence of the perturbations $e^{-i\omega t}$ is assumed and a
$z$-dependence of the form $e^{i k_z z}$ is taken, with $k_z= \pi/L$ to
incorporate the line-tying effect. The governing equations of our system reduce
to a scalar Helmholtz equation for the magnetic pressure. This is solved
analytically with the $T$-matrix theory \citep[see,
e.g.,][]{bogdan&cattaneo1989,keppens1994,luna2009}. 

In order to compare the dynamics of a multi-stranded loop model with that of a 
monolithic tube, an equivalent flux tube is defined. The flux tube radius, $R$,
corresponds to that of the cylinder that wraps the strand bundle (see Figure
\ref{sketch_multistrand}). The equivalent uniform density is
\begin{equation}\label{dens_eq}
\rho_\mathrm{eq}=\sum_{j=1}^{N} \rho_{j} \left(
\frac{a}{R} \right)^2 + \rho_\mathrm{e} \left[ 1 - N
\left(\frac{a}{R}\right)^2 \right]~,
\end{equation}
where the mass of the strand set and the coronal medium inside the hypothetical
monolithic loop are considered. We have fixed the radius of the cylinder envelop
to $R=0.03 L$, a typical value for coronal loops \citep[see][]{aschwanden2003}.
We have assumed the volume filled by the strands to be $40\%$ that of the
monolithic loop. In addition, all the strands have the same radius. In this
work, we have considered systems of $10$ and $40$ strands with radii $a=0.2 R=
0.006 L$ and $a=0.1 R= 0.003 L$, respectively.

\section{Normal modes of ten identical strands}\label{10id_strands}

We first study a system of $N=10$ identical strands, i.e., with identical
densities and radii. From the results of \citet{luna2009}, this is
the situation for which the coupling between strands is stronger because all the
tubes have identical individual kink frequencies, hereafter denoted by
$\omega_\mathrm{strand}$. The density
of each strand is fixed to $\rho_{j}=7.5\rho_\mathrm{e}$, which yields the
equivalent
density $\rho_\mathrm{eq}=3.6\rho_\mathrm{e}$ (see Equation (\ref{dens_eq})).
The equivalent
monolithic loop has an individual kink frequency $\omega_\mathrm{mono}=
2.067v_\mathrm{A e}/L$
computed with the fast wave dispersion relation in a cylinder
\citep{edwin&roberts1983}, with $v_\mathrm{A e}$ the Alf\'en speed in the
coronal environment. Hereafter, all the frequencies are expressed in terms of
this frequency. The individual kink frequency of each strand is then
$\omega_\mathrm{strand}=0.737\omega_\mathrm{mono}$.

\subsection{Frequency analysis of the collective normal modes}

\begin{figure}[!ht]
\centering
\mbox{\hspace{-0.cm}\hspace{11.cm}}\vspace{-2.cm}
\includegraphics[width=11.cm]{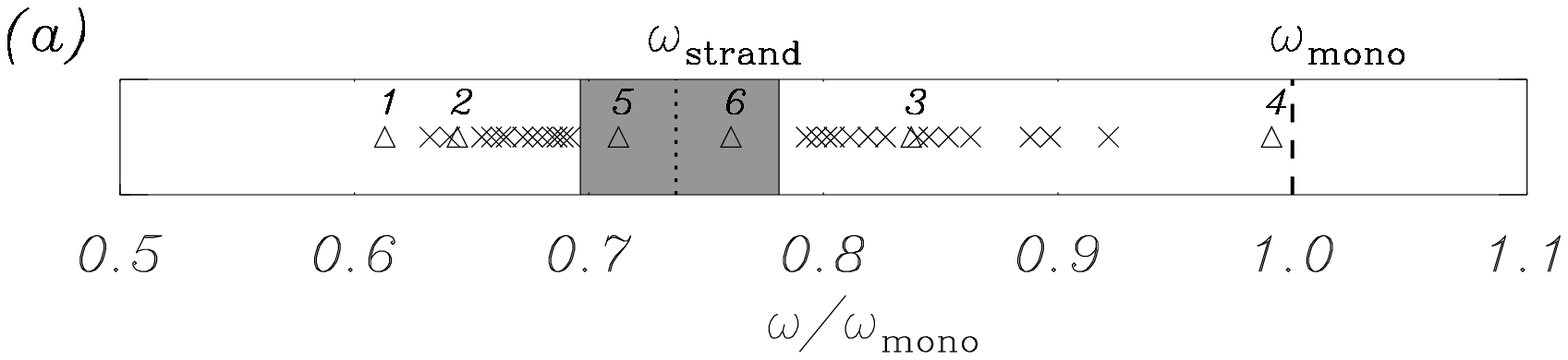}\vspace{-3.cm}
\mbox{\hspace{-0.cm}\hspace{11.cm}}\vspace{-2.cm}
\includegraphics[width=11.cm]{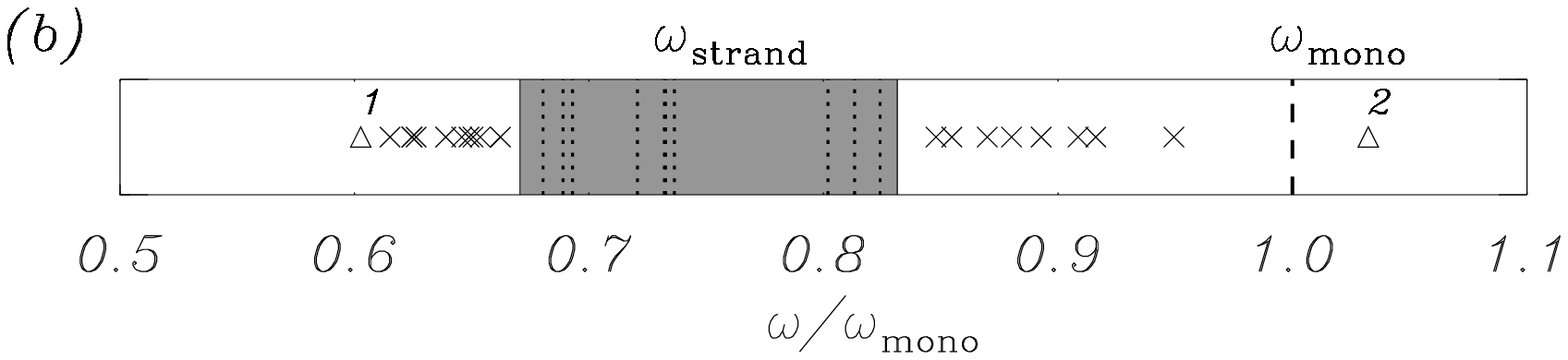}\vspace{-3.cm}
\mbox{\hspace{-0.cm}\hspace{11cm}}\vspace{-2.cm}
\includegraphics[width=11.cm]{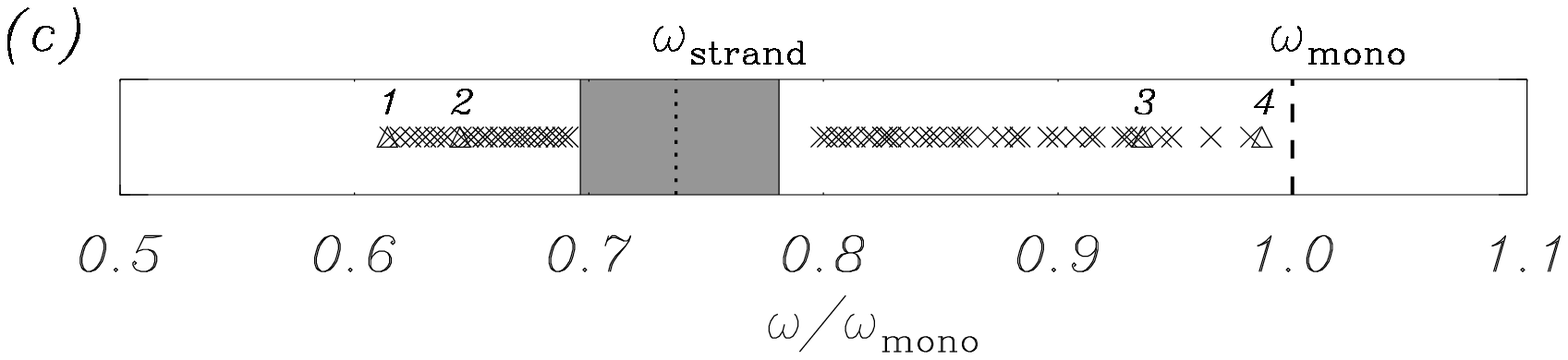}\vspace{-2.cm}
\caption{Frequency distribution of the collective normal modes associated to the
three systems considered in this work: {\bf (a)} $10$ identical strands, {\bf
(b)} $10$ nonidentical strands with different densities, and {\bf (c)} $40$
identical strands. In all cases we clearly see that the frequencies are
distributed at both sides of the individual strand frequencies,
$\omega_\mathrm{strand}$, (dotted line) in a broad band of frequencies and that
all modes have frequencies mainly below $\omega_\mathrm{mono}$ (dashed line).
A shaded area is plotted between the lowest and largest mid mode
frequencies. Then, mid modes lie within the shaded area, whereas low and high
modes lie to its left and right, respectively.
The triangles mark the frequencies of the modes whose spatial structure is
displayed in the following plots and are labeled with integers.}
\label{eigenfrequencies}
\end{figure}

We have investigated the eigenfrequencies of the system and have found that they
are distributed at both sides of the individual strand frequency and always
below the frequency of the equivalent monolithic loop (see Figure
\ref{eigenfrequencies}(a)). The lowest and highest frequencies are
$\omega=0.612\omega_\mathrm{mono}$ and $\omega=0.993\omega_\mathrm{mono}$,
respectively. We
see that the eigenfrequencies are in a broad band of width approximately
$0.38\omega_\mathrm{mono}$. According to their spatial structure, we classify
the
normal modes in three groups. Modes with frequencies below the central frequency
($\omega\lesssim\omega_\mathrm{strand}$) are called low modes (left-hand side of
the shaded
area in Figure \ref{eigenfrequencies}(a)). Mid modes are those with frequencies
similar to the central frequency ($\omega \approx \omega_\mathrm{strand}$;
shaded area
in Figure \ref{eigenfrequencies}(a)), and finally the solutions with $\omega
\gtrsim
\omega_\mathrm{strand}$ are referred to as high modes (right-hand side of the
shaded
area in Figure \ref{eigenfrequencies}(a)). It is important to note that in a
system of non-interacting strands the frequency of oscillation of each strand is
$\omega_\mathrm{strand}$.

\subsection{Velocity and total pressure perturbation analysis}

\begin{figure}[!ht]
\centering
\includegraphics[width=8.cm]{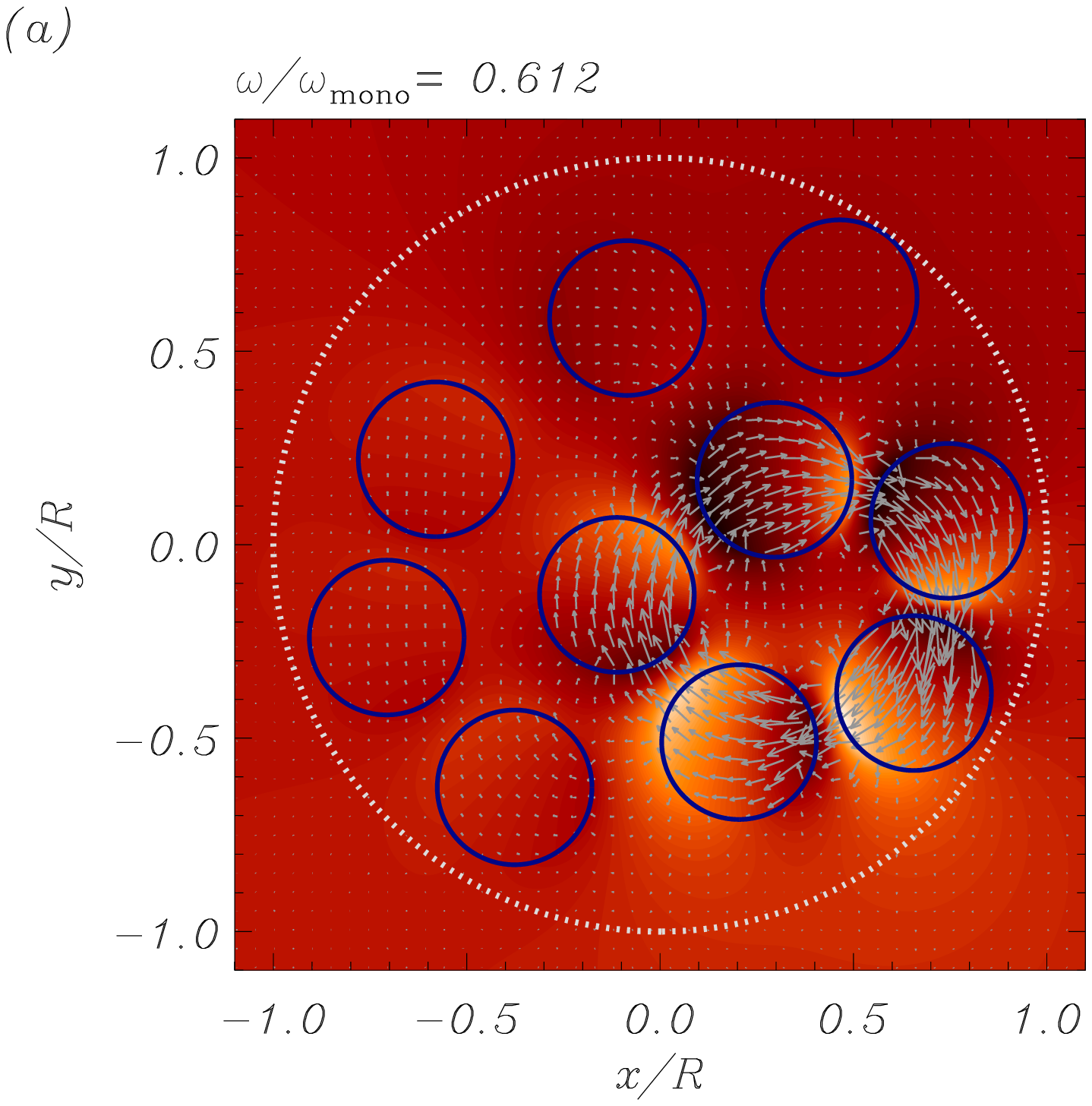}\hspace{-0.cm}\includegraphics[
width=8.cm]{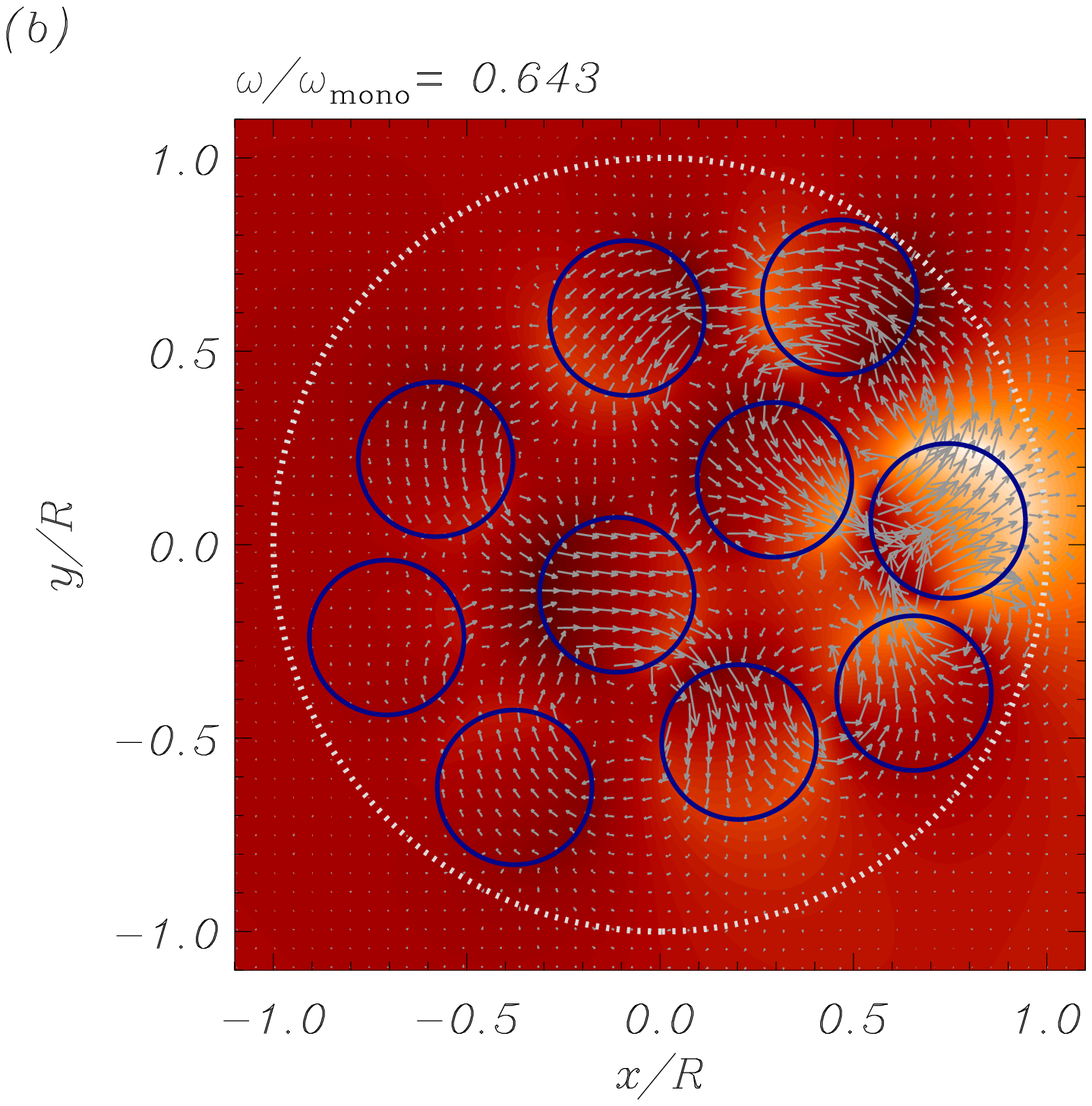}\vspace{-2.cm}
\includegraphics[width=8.cm]{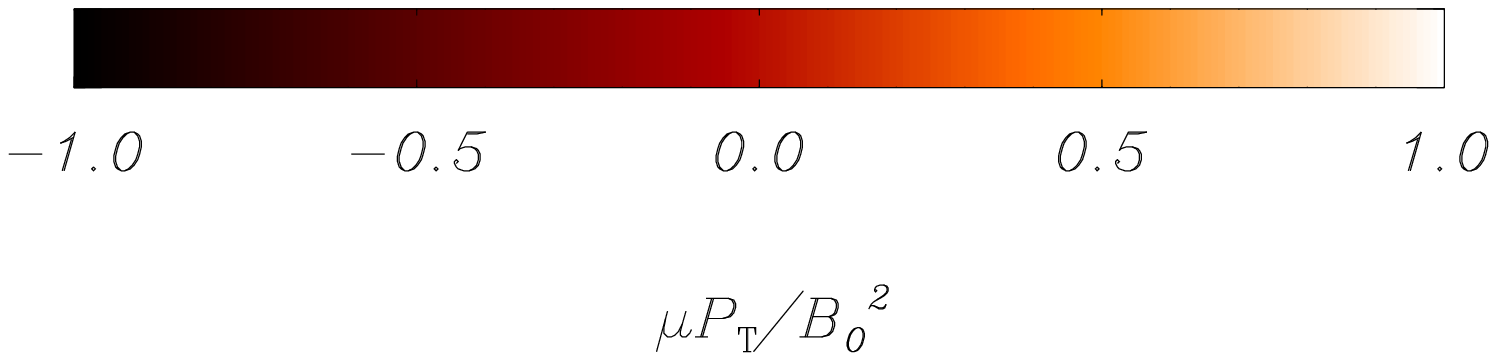}\vspace{-1.cm}
\caption{Total pressure perturbation (color field) and velocity field (arrows)
of the fast collective normal modes of the two low modes labeled as $1$ and $2$
in Figure \ref{eigenfrequencies}(a). {\bf(a)} Lowest frequency mode labeled as
$1$. {\bf(b)} Low mode, labeled as $2$.}
\label{bottom_modes}
\end{figure}

The spatial structure of the three groups of modes is clearly different. Low
modes are kink-like modes in the sense that at least one strand moves
transversely like in a kink mode of an individual loop. For these modes, the
fluid between tubes follows the strand motion (see Figure \ref{bottom_modes}),
producing chains of loops in which one follows the next. In Figure
\ref{bottom_modes}, two examples of low modes are plotted. Figure
\ref{bottom_modes}(a) corresponds to the lowest frequency mode, in which only
five strands oscillate, producing some kind of global torsional motion of the
strands. In Figure \ref{bottom_modes}(b) another example of low eigenfunction is
plotted and it shows that almost all the strands are excited. As in the previous
example, the fluid between strands moves with them. In both modes the maximum
velocity takes place inside the strands. These characteristics are shared by all
the low modes. The $S_x$ and $A_y$ modes of the system of two loops of
\citet{luna2008} and the $m_{1}$ to $m_{4}$ modes of a system of three aligned
loops of \citet{luna2009} can be classified in the low-mode group because the
spatial structure of the magnetic pressure perturbation and velocity fields have
the features previously described and their frequencies are below the
corresponding individual kink frequency.

\begin{figure}[!ht]
\centering
\includegraphics[width=8.cm]{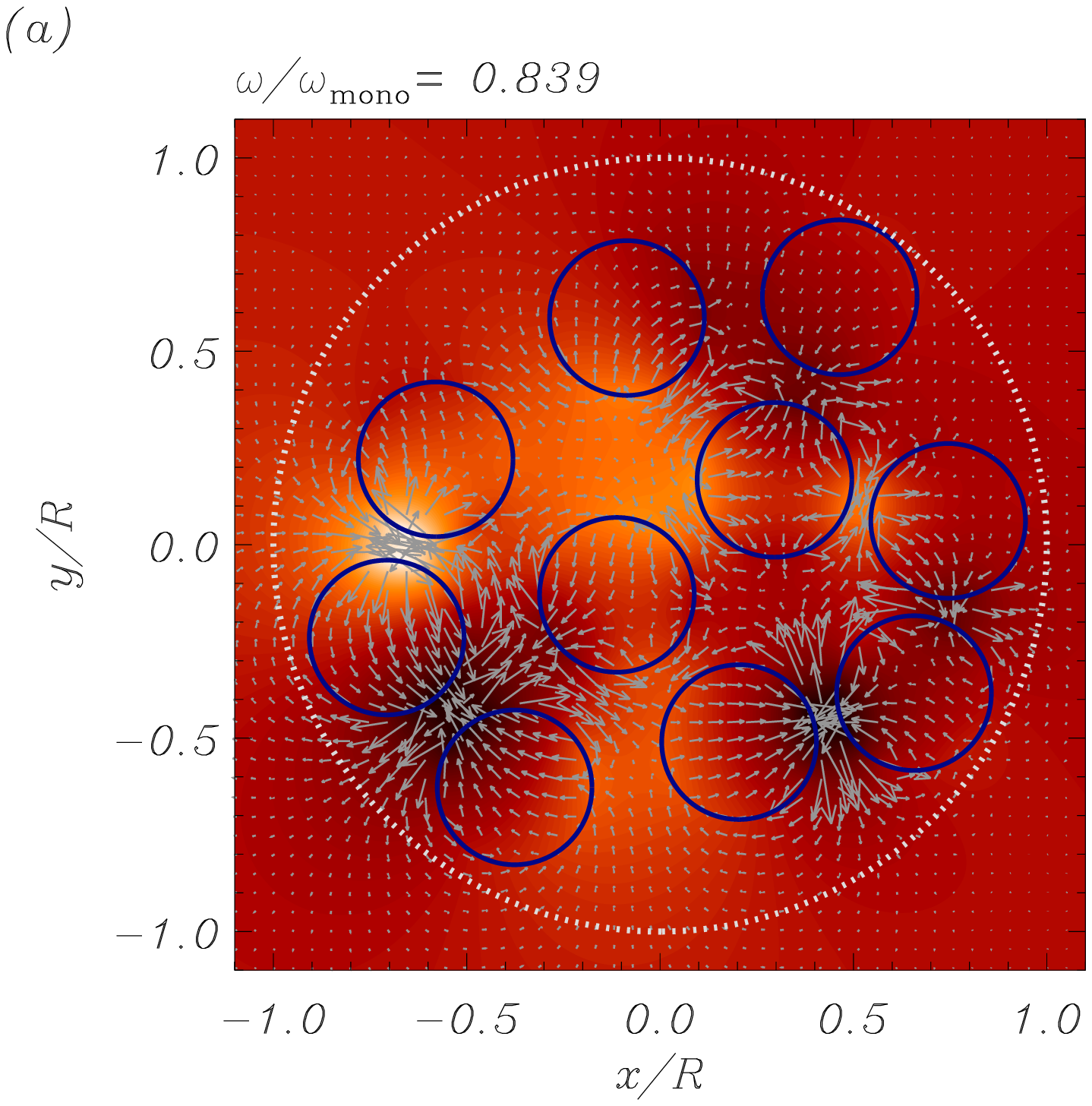}\hspace{-0.cm}\includegraphics[
width=8.cm]{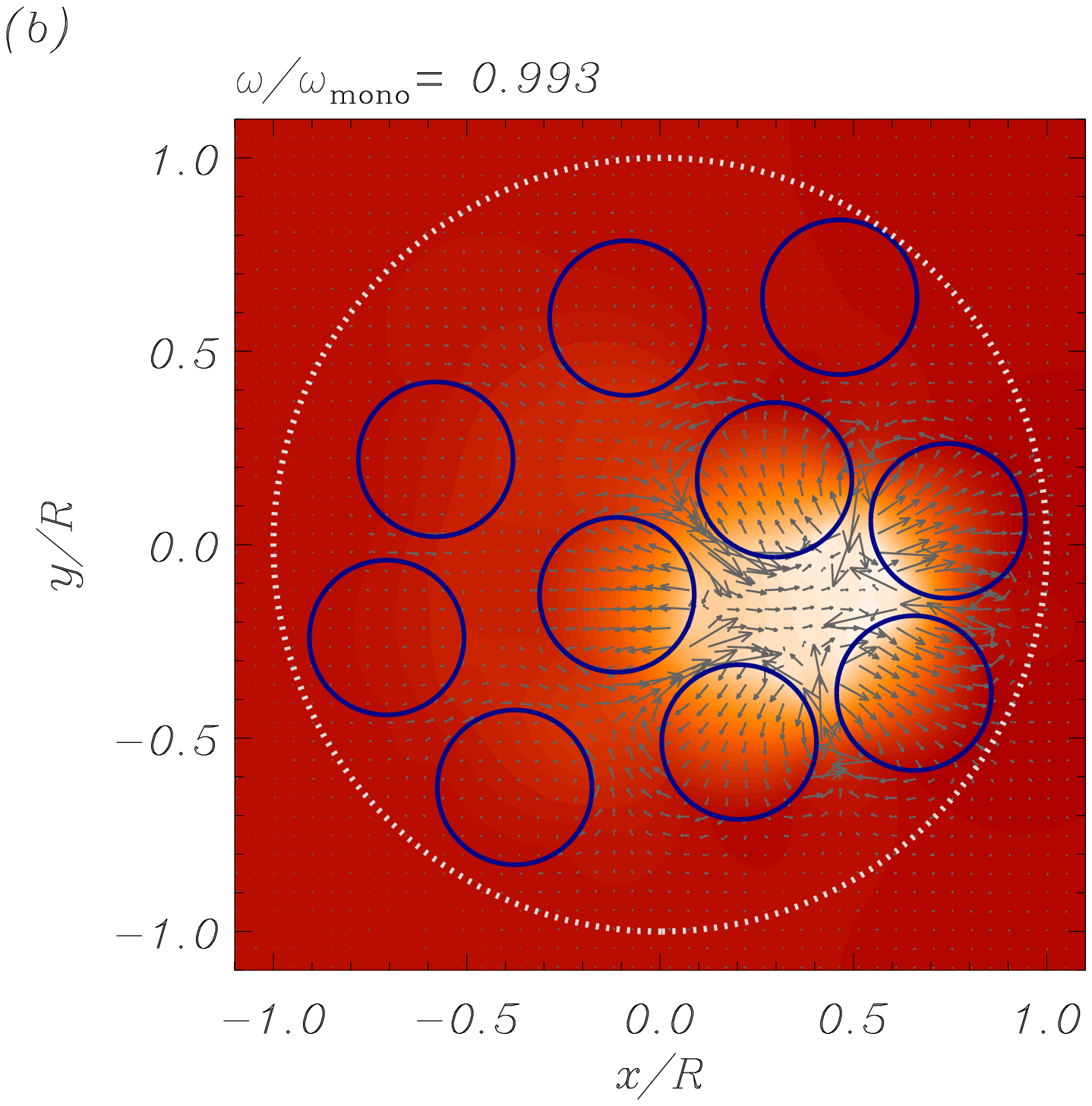}\vspace{-2.cm}
\includegraphics[width=8.cm]{bar.eps}\vspace{-1.cm}
\caption{Same as Figure \ref{bottom_modes} for two high modes. {\bf (a)} Mode
labeled as $3$ in Figure \ref{eigenfrequencies}(a). {\bf(b)} Highest frequency
mode, labeled as $4$.}
\label{top_modes}
\end{figure}

On the other hand, for the high modes (see Figure \ref{top_modes}) the
intermediate fluid between tubes is compressed or rarefied (which leads to a
higher or lower total pressure perturbation) or moves in the opposite direction
to the strands, producing a more forced motion than that of the low modes. High
modes are kink-like too, but in contrast to the low modes the maximum velocities
take place in the intermediate fluid between strands. This behavior is very
clear in Figure \ref{top_modes}(a), in which the strand motions force the
coronal fluid to pass through the narrow channels between them or to compress
the coronal medium. Similarly, in the highest frequency mode (Figure
\ref{top_modes}(b)), high velocity flows between the five excited strands takes
place. The coronal medium within the excited strands is compressed and rarefied,
giving rise to some kind of sausage global motion of the strands. All the modes
that we have classified as high share these characteristics. The $S_y$ and $A_x$
modes of two identical tubes of \cite{luna2008} and the $m_{5}$ to $m_{8}$ modes
of a system of three aligned loops of \cite{luna2008} belong to the high-mode
group.

\begin{figure}[!ht]
\centering
\includegraphics[width=8.cm]{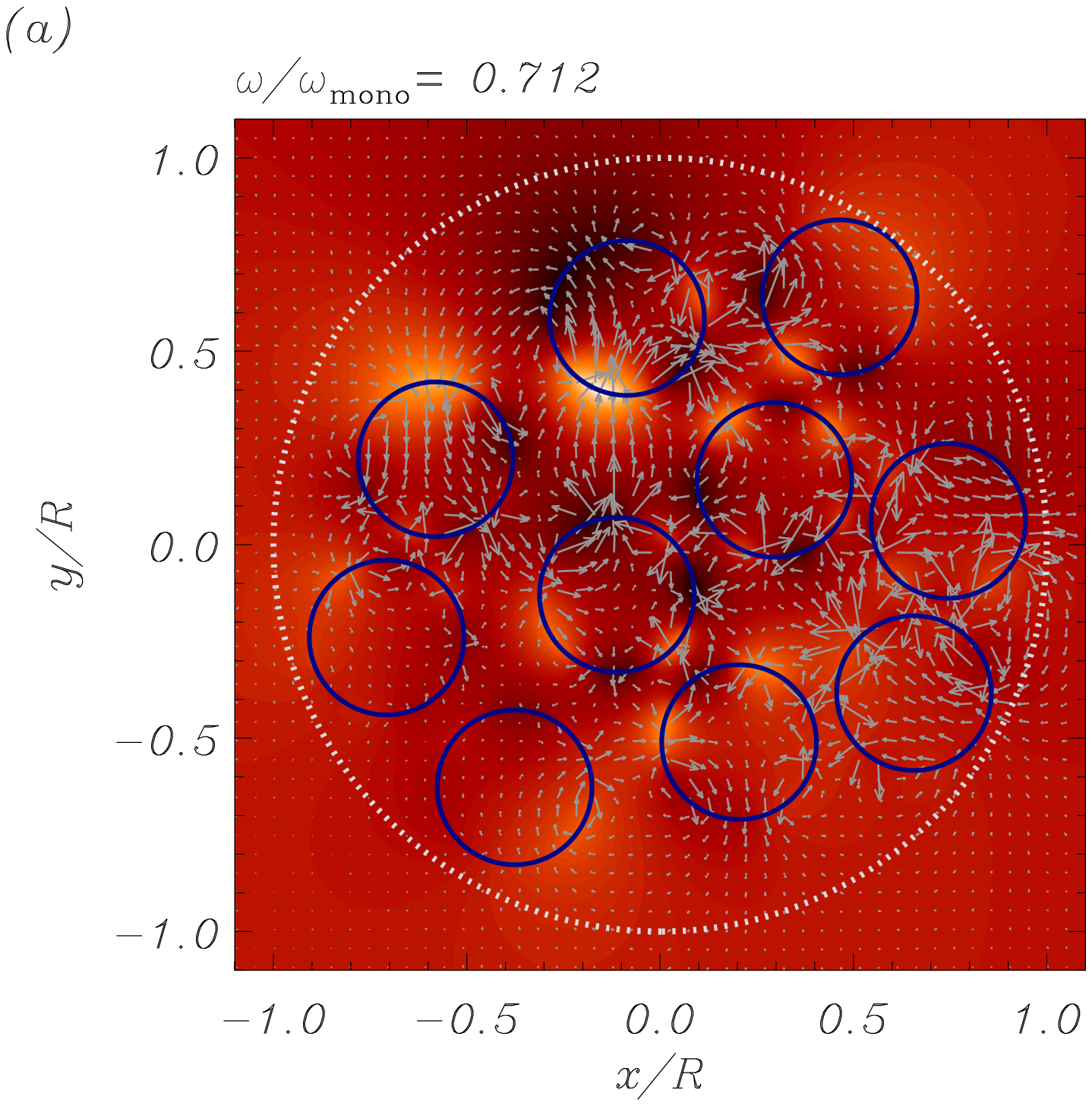}\hspace{-0.cm}\includegraphics[
width=8.cm]{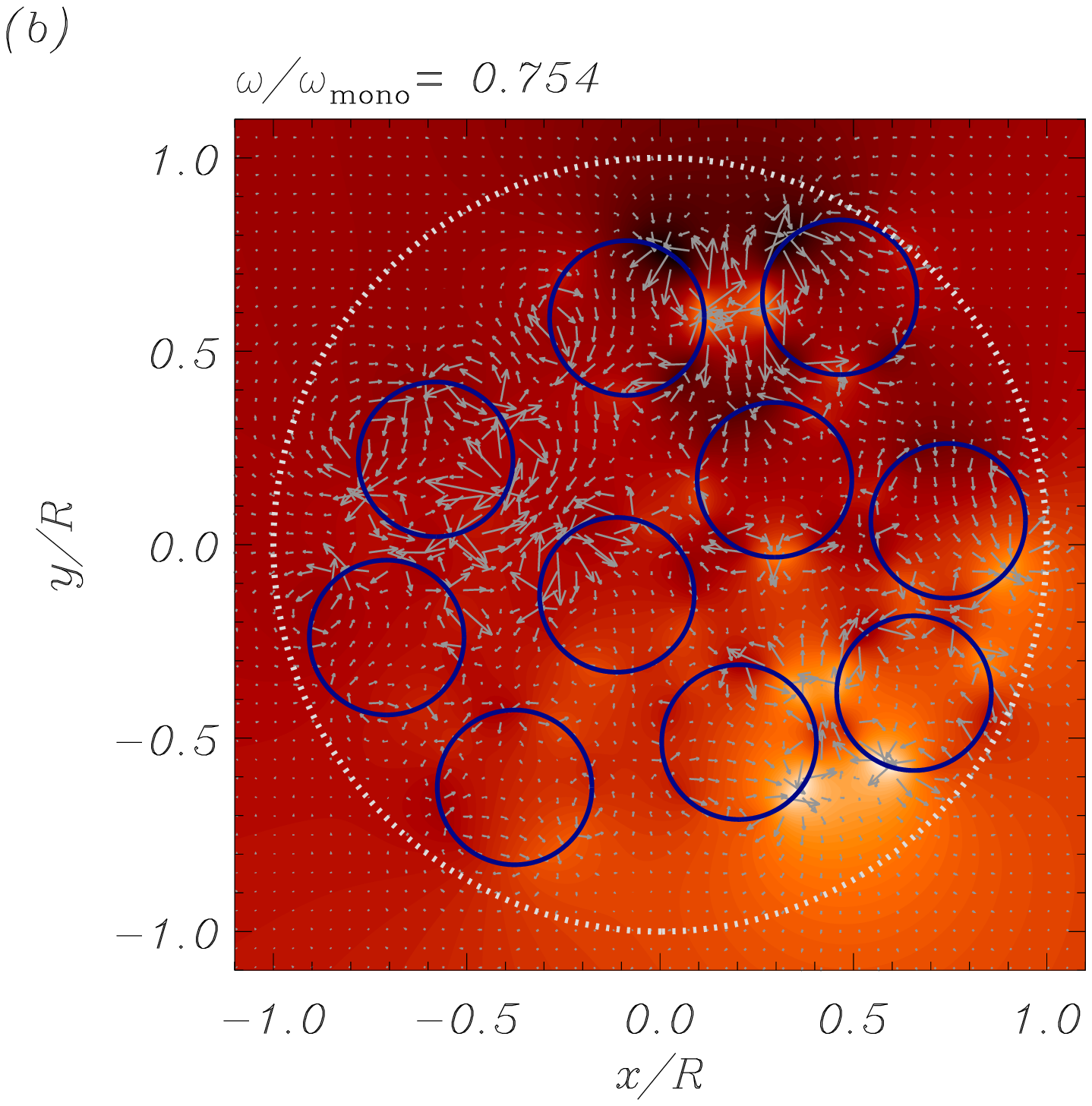}\vspace{-2.cm}
\includegraphics[width=8cm]{bar.eps}\vspace{-1.cm}
\caption{Same as Figure \ref{bottom_modes} for two mid modes. {\bf(a)} Mode
labeled as $5$ in Figure \ref{eigenfrequencies}(a). {\bf(b)} Mode labeled as
$6$. In both cases we see the complex structure of the mid normal modes.}
\label{middle_modes}
\end{figure}

Finally, the mid modes have the most complex spatial structure. They are
fluting-like modes and have strand motions similar to those of the fluting modes
of an individual tube (see Figure \ref{middle_modes}). The magnetic pressure
perturbation and velocity are concentrated mainly in the strand surface. There
is an infinite number of mid modes with frequencies concentrated around $\omega
\approx \omega_\mathrm{strand}$, and for this reason they are plotted as a
shaded area in Figure \ref{eigenfrequencies}.

\section{Normal modes of ten non-identical strands}\label{non_id_strands}

\begin{figure}[!ht]
\centering
\includegraphics[width=8.cm]{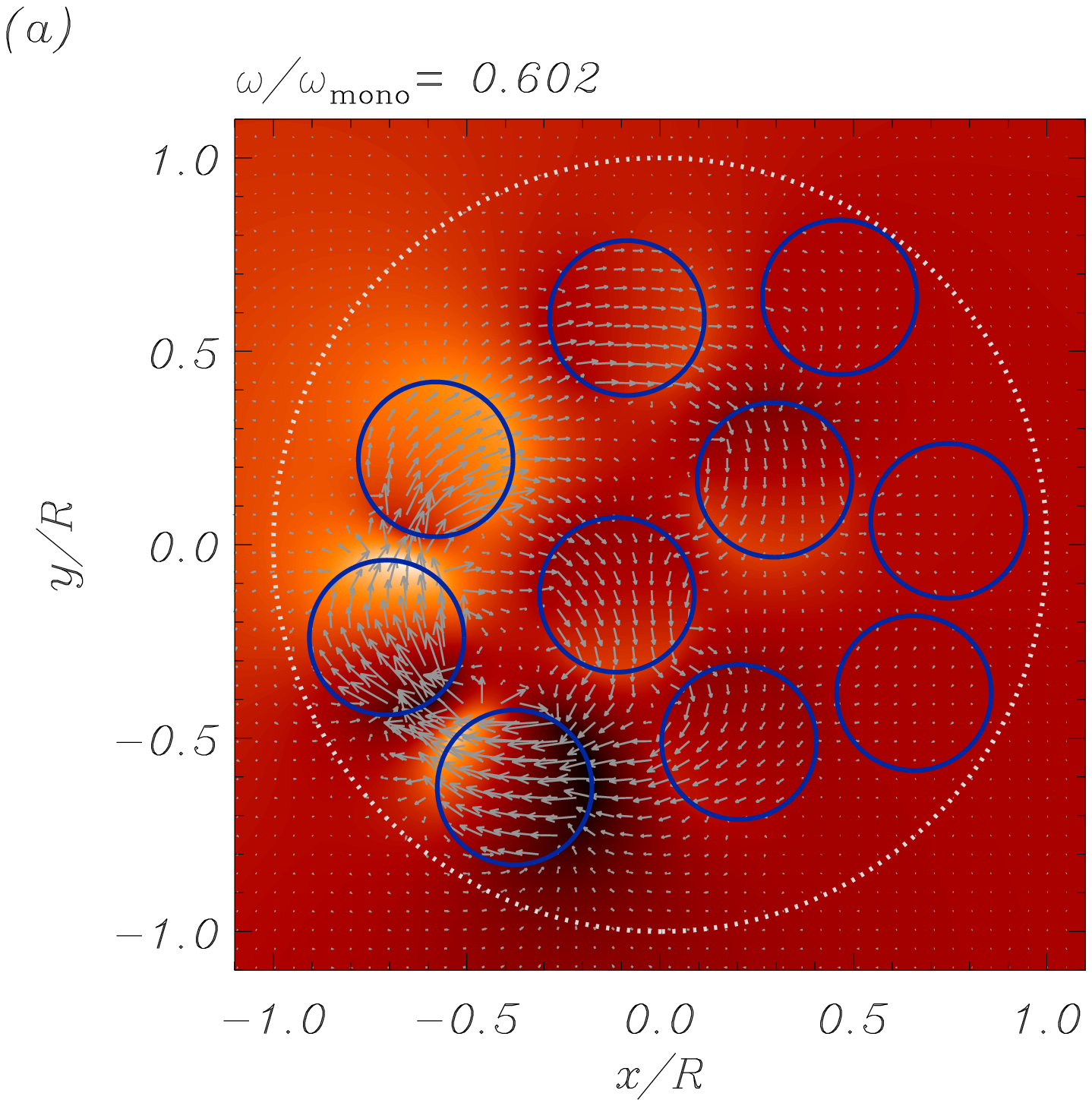}\hspace{-0.cm}\includegraphics[
width=8.cm]{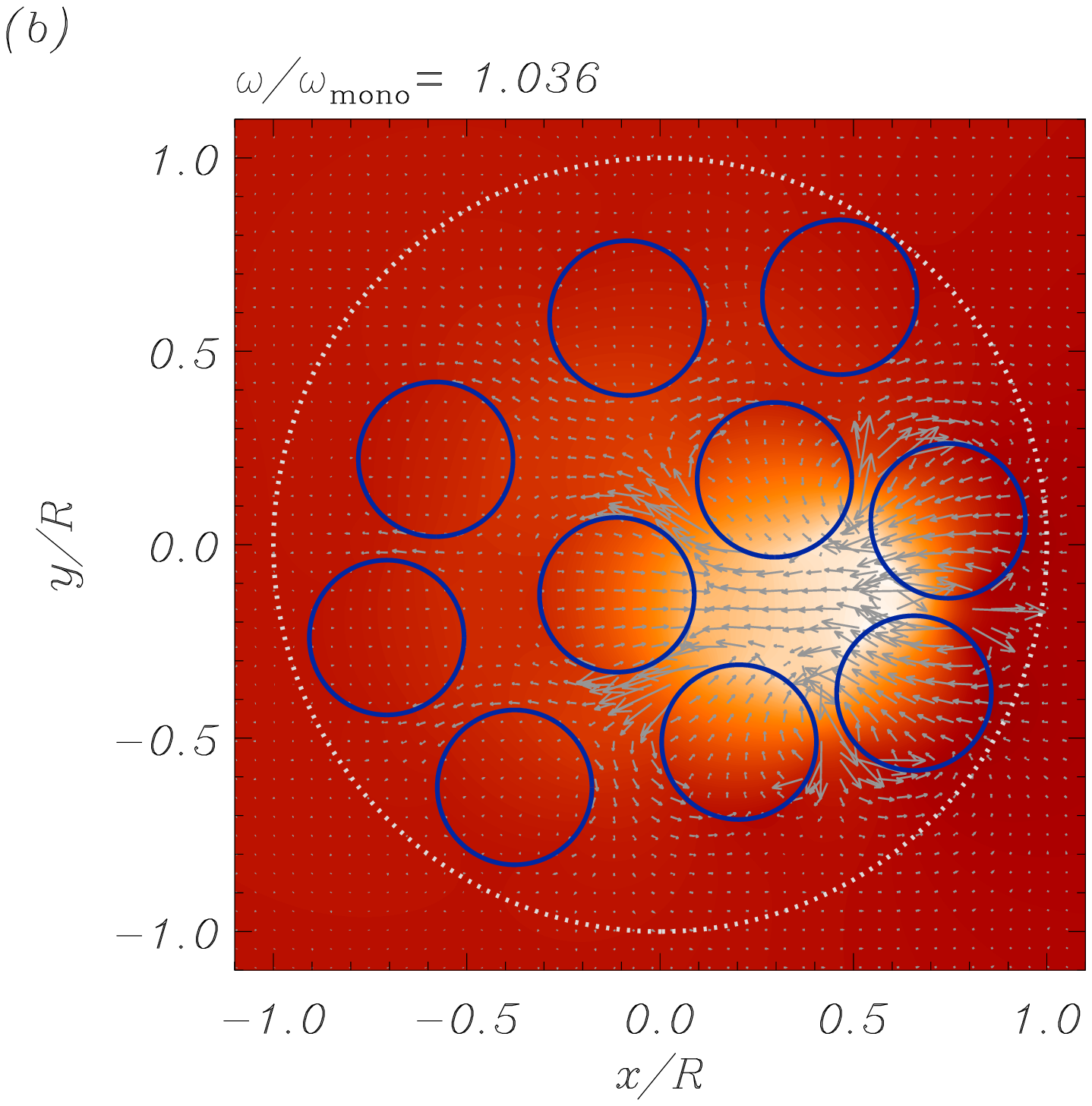}\vspace{-2.cm}
\includegraphics[width=8.cm]{bar.eps}\vspace{-1.cm}
\caption{Same as Figure \ref{bottom_modes} for the collective normal modes of a
system of $10$ non-identical strands. {\bf (a)} Lowest frequency mode, labeled
as $1$ in Figure \ref{eigenfrequencies}(b). {\bf (b)} Highest frequency mode,
labeled as $2$.}
\label{modes_nonid}
\end{figure}

In this Section we have considered the previous spatial distribution of strands
but with different densities. The strand densities have been distributed
randomly around and average density $7.5\rho_\mathrm{e}$ within a range of
$3\rho_\mathrm{e}$. The equivalent monolithic density has been kept equal to
$\rho_\mathrm{eq}=3.6\rho_\mathrm{e}$ and the volume filled by the strands to
$40\%$ that of the monolithic loop volume, as in \S \ref{10id_strands}. The
densities we use are $\rho_{j}/\rho_\mathrm{e}=\{7.89$, $7.61$, $7.60$, $8.97$,
$5.98$, $8.73$, $7.52$, $8.62$, $6.18$, $5.80\} $ following the ordination of
Figure \ref{sketch_multistrand}. The considered range of strand densities
implies that the difference between the maximum and minimum values of the
individual kink frequencies is $0.13\omega_\mathrm{mono}$. This  makes the
coupling between the strands weaker \citep[see][]{luna2009} than in the
identical strand case discussed in \S \ref{10id_strands}. However, the strands
still interact and so it is not possible to consider the multi-stranded system
as a collection of individual tubes. The band of collective frequencies now goes
from $\omega=0.602\omega_\mathrm{mono}$ to $\omega=1.036\omega_\mathrm{mono}$,
i.e., it has a width of $0.43\omega_\mathrm{mono}$, as we see in Figure
\ref{eigenfrequencies}(b). This band is broader than in the identical strand
case (for which it is $0.38\omega_\mathrm{mono}$), but this does not mean that
the interaction between non-identical strands is stronger. The reason is the
additional broadening associated to the spreading of the individual kink
frequencies, which results in the enlargement of the mid frequency band (see
Figure \ref{eigenfrequencies}(b)). Roughly speaking, the broadening associated
to the coupling is then the total broadening minus the spreading of the
individual kink frequencies. In case of an uncoupled system of nonidentical
strands the width of the band associated to the coupling is zero. The individual
kink frequencies of our system are in a band of $0.13\omega_\mathrm{mono}$. This
implies that the contribution of the strand interaction is roughly
$0.30\omega_\mathrm{mono}$, indicating less interaction between the strands than
for the identical strand system  of \S \ref{10id_strands}. Similarly to \S
\ref{10id_strands}, we can divide the collective normal modes in three groups
(low, mid, and high). However, the spatial structure differs from those of the
previous Section. The differences are clear, for example, in the lowest
frequency mode. Comparing Figure \ref{modes_nonid}(a) with Figure
\ref{bottom_modes}(a) we see that the global torsional oscillation of the five
strands labeled $2$, $5$, $6$, $7$, and $10$ is avoided because their densities
are very different, but the oscillation of the strands labeled as $1$, $2$, $3$,
$4$, $6$, $7$, and $8$ with similar densities, is favored. The highest frequency
mode plotted in Figure \ref{modes_nonid}(b) is very similar to the corresponding
mode in the identical tube case (Figure \ref{top_modes}(b)), although the
amplitude of the oscillations is concentrated in the rarest tubes, labeled $5$
and $10$. These results are general and so low modes have the largest
oscillatory amplitudes in the denser tubes. On the contrary, for the high modes,
the highest oscillatory amplitudes are associated to the rarest strands. Mid
modes have a complex spatial structure but similar to that of the identical
strand case and are not plotted for the sake of simplicity.

In \citet{terradas2008} a system of $10$ non-homogeneous strands was considered.
The authors studied the time-dependent evolution of the system after an initial
excitation. They found a collective frequency $0.22/\tau_\mathrm{A}$, where
$\tau_\mathrm{A}$ is defined as $\tau_\mathrm{A}=R/v_\mathrm{A e}$. We have
considered an equivalent system of homogeneous strands preserving the total mass
and have found that modes lie in a frequency band going from
$0.182/\tau_\mathrm{A}$ to $0.23/\tau_\mathrm{A}$ that agrees very well with the
mentioned results.

\section{Normal modes of forty identical strands}\label{more_complex}

\begin{figure}[!ht]
\centering
\includegraphics[width=8.cm]{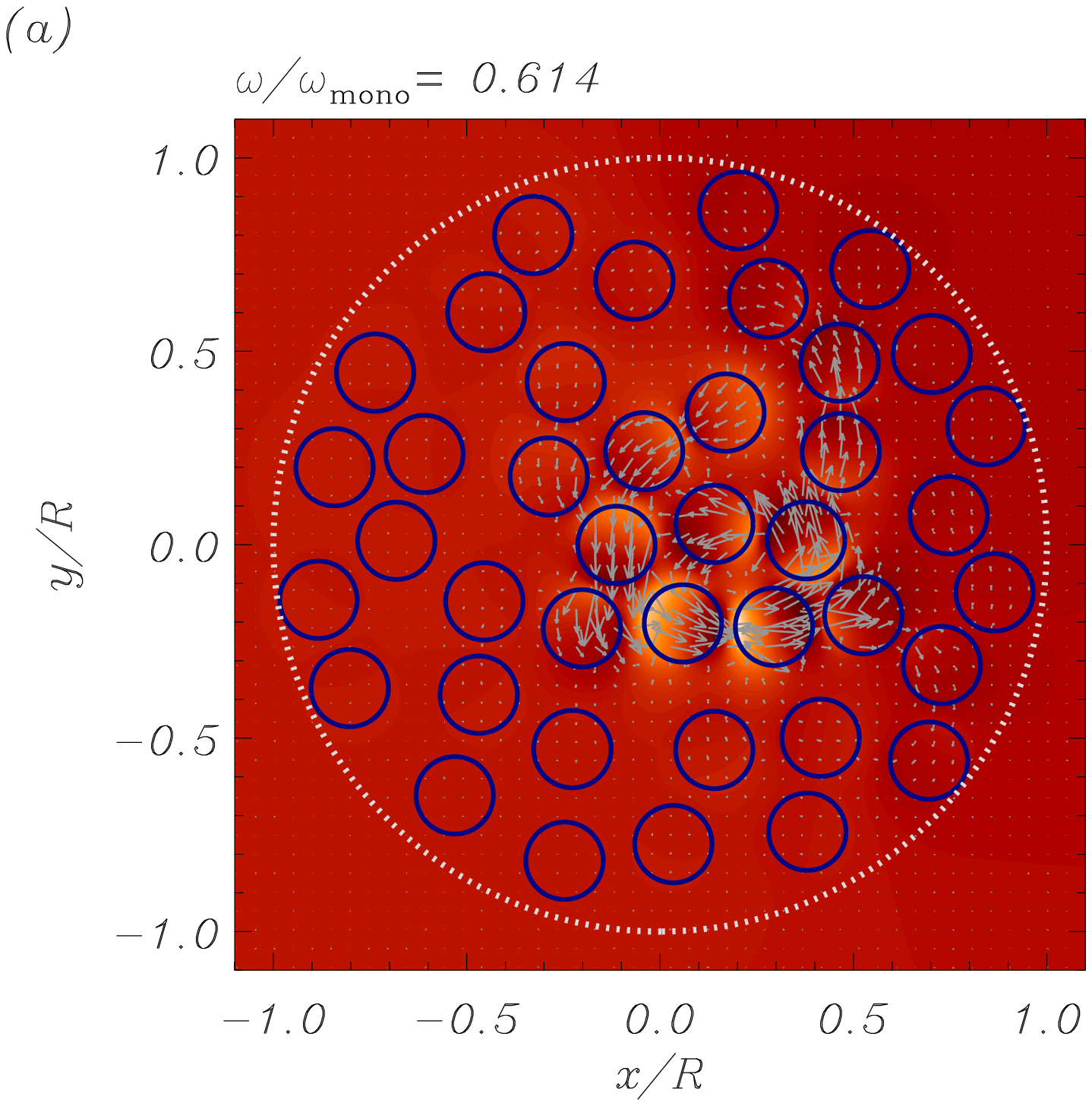}\hspace{-0.cm}\includegraphics[
width=8.cm]{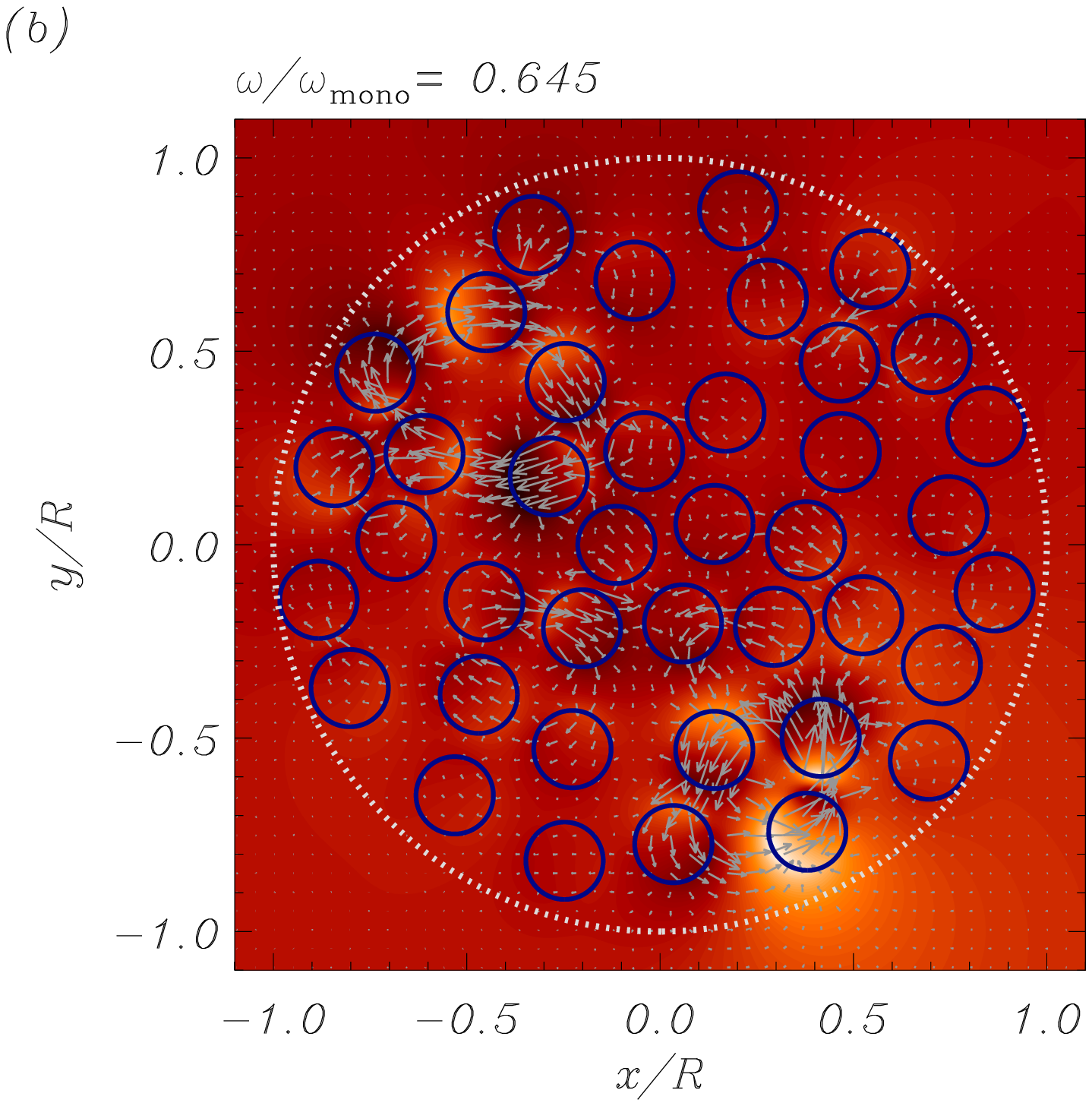}\vspace{-2.cm}
\includegraphics[width=8.cm]{bar.eps}\vspace{-1.cm}
\caption{Same as Figure \ref{bottom_modes} for two low modes in a system of $40$
identical strands. {\bf(a)} Lowest frequency mode, labeled as $1$ in Figure
\ref{eigenfrequencies}(c). {\bf (b)} Normal mode labeled as $2$.}
\label{bottom_n40}
\end{figure}

\begin{figure}[!ht]
\centering
\includegraphics[width=8.cm]{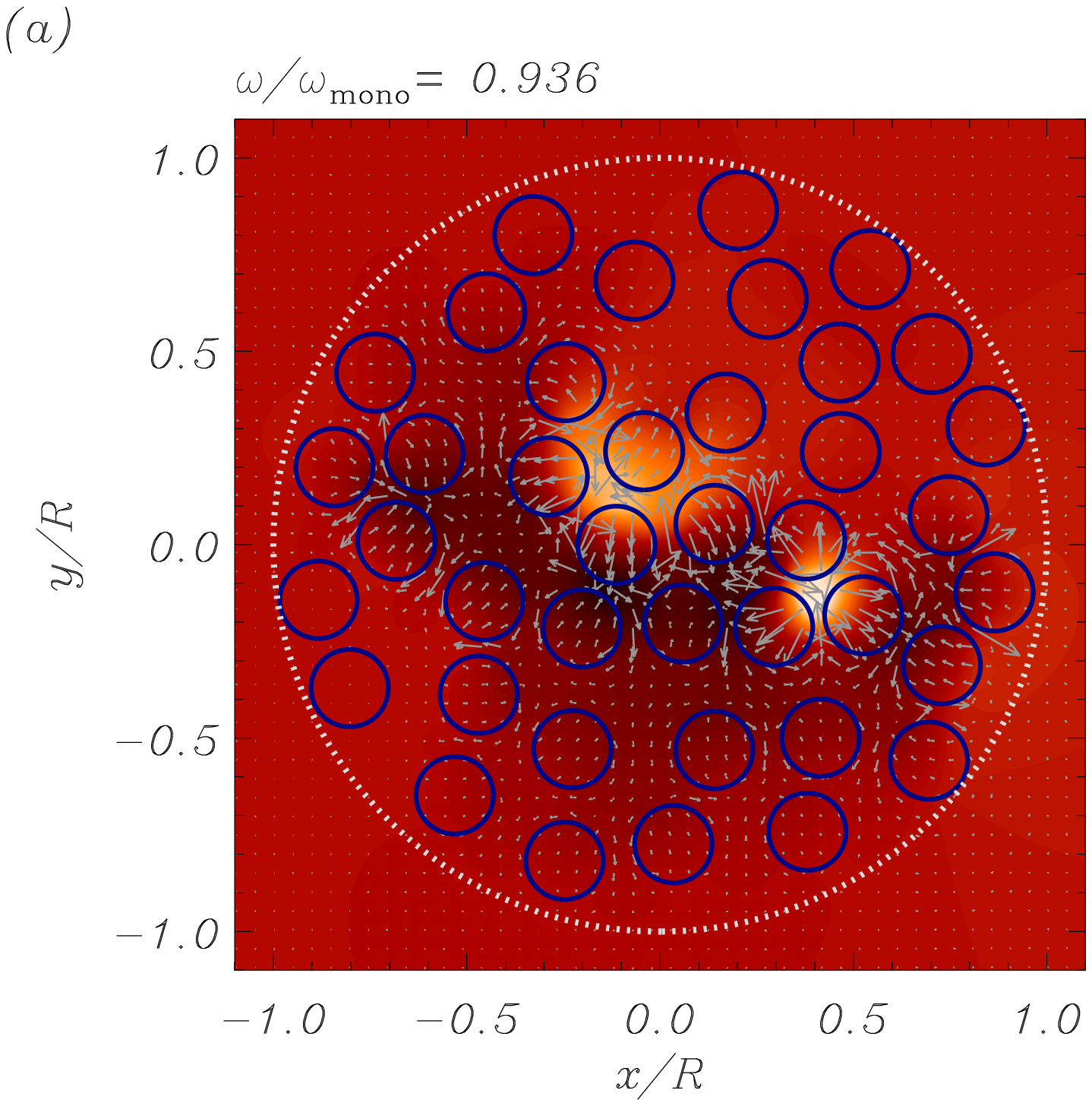}\hspace{-0.cm}\includegraphics[
width=8.cm]{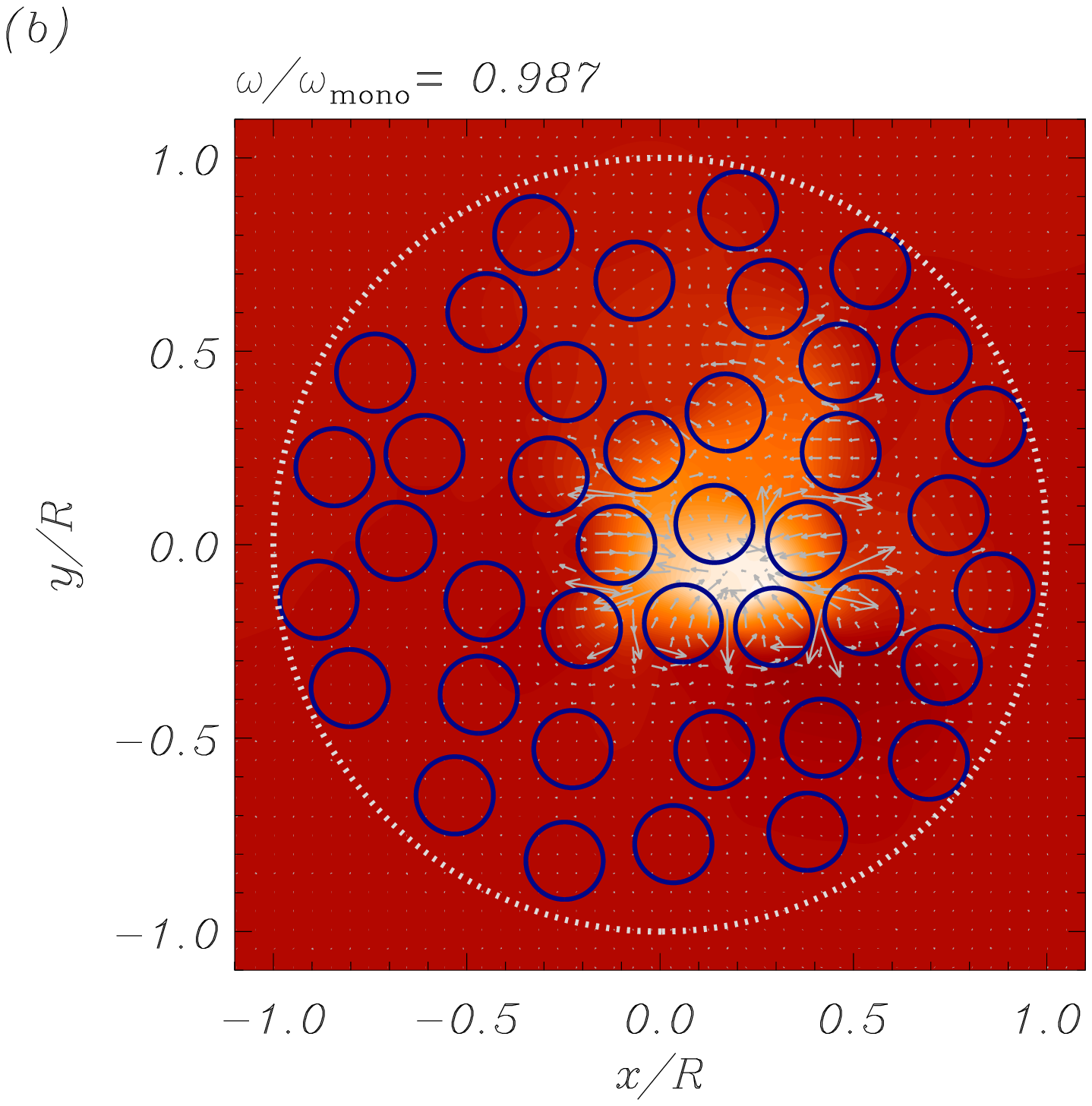}\vspace{-2.cm}
\includegraphics[width=8.cm]{bar.eps}\vspace{-1.cm}
\caption{Same as Figure \ref{bottom_modes} for two high modes in a system of
$40$ identical strands. {\bf(a)} Collective normal mode, labeled as $3$ in
Figure \ref{eigenfrequencies}(c). {\bf(b)} Highest frequency mode, labeled as
$4$.}
\label{top_n40}
\end{figure}

We have also investigated the normal modes of a much more complex system of $40$
identical strands. The strands fill $40\%$ of the equivalent loop volume, with a
strand density $\rho_j=7.5\rho_\mathrm{e}$ and an equivalent density
$\rho_\mathrm{eq}=3.6\rho_\mathrm{e}$. The frequencies of the normal modes lie
in a band that goes from $\omega=0.614\omega_\mathrm{mono}$ to
$\omega=0.987\omega_\mathrm{mono}$, so that its width is
$0.37\omega_\mathrm{mono}$. This frequency band is similar to that of the $10$
identical strand case (see Figures \ref{eigenfrequencies}(a) and
\ref{eigenfrequencies}(c)). However, the system of $40$ strands has more
collective normal modes than the system of $10$ strands. The classification in
low, mid, and high modes is still valid in this complex system of strands. In
this Section we have only considered the kink-like modes (low and high modes)
and the mid modes are not plotted for the sake of simplicity. In Figures
\ref{bottom_n40}(a) and \ref{bottom_n40}(b), two examples of low collective
normal modes are plotted. In the lowest frequency normal mode (Figure
\ref{bottom_n40}(a)), a cluster of close strands is excited and the others are
at rest. In the second example (Figure \ref{bottom_n40}(b)), a cluster of
distant strands participate in the motion. In Figures \ref{top_n40}(a) and
\ref{top_n40}(b), two examples of high modes are also plotted. Similarly to the
low modes, in the high modes a cluster of several strands participates in the
motion whereas the others are at rest.

\section{Time-dependent analysis: numerical simulations}\label{time_evolution}

In the previous sections we have considered the normal modes of a multi-stranded
loop. Normal mode analysis provides information about the stationary state of
the system or ideally, at an infinite time. However, loop oscillations are often
produced by an impulsive event like a flare and it is more suitable to describe
such events in terms of an initial value problem \citep[see,
e.g.,][]{terradas2009}. In addition, the time-dependent analysis gives
information on how the different collective normal modes are excited and on how
they are related with the temporal evolution after the initial disturbance.

We shall consider here the temporal evolution of the multi-stranded loop
composed of $10$ identical strands of \S \ref{10id_strands}. The governing
equations of the temporal evolution of the velocity field, $\mathbf{v}=\left(
v_x, v_y,0 \right)$, and magnetic field perturbation, $\mathbf{B} =\left( B_x,
B_y, B_z \right)$, are the linearized ideal MHD equations, namely
\begin{eqnarray}\label{lmhd_linetying1}
\frac{\partial v_x}{\partial t} &=& \frac{v_\mathrm{A}^2}{B_{0}} \left( k_z ~
\widetilde{B}_x - \frac{\partial B_z}{\partial x} \right)~,
\\\label{lmhd_linetying2}
\frac{\partial v_y}{\partial t} &=& \frac{v_\mathrm{A}^2}{B_0} \left( k_z ~
\widetilde{B}_y - \frac{\partial B_z}{\partial y} \right)~,
\\\label{lmhd_linetying3}
\frac{\partial \widetilde{B}_x}{\partial t} &=& -~B_0~k_z~v_x ~,
\\\label{lmhd_linetying4}
\frac{\partial \widetilde{B}_y}{\partial t} &=& -~B_0~k_z~v_y ~,
\\\label{lmhd_linetying5}
\frac{\partial B_z}{\partial t} &=& -B_0 \left( \frac{\partial v_x}{\partial x}
+ \frac{\partial v_y}{\partial y} \right)~, 
\end{eqnarray}
where $B_x=-i \widetilde{B}_x$ and $B_y=-i \widetilde{B}_y$ are purely imaginary
variables. This fact indicates that the $x$- and $y$-components of the magnetic
field have a phase lag of $\pm \pi/2$ with respect to the temporal evolution of
the other variables. 

\begin{figure}[!ht]
\centering
\includegraphics[width=7.cm]{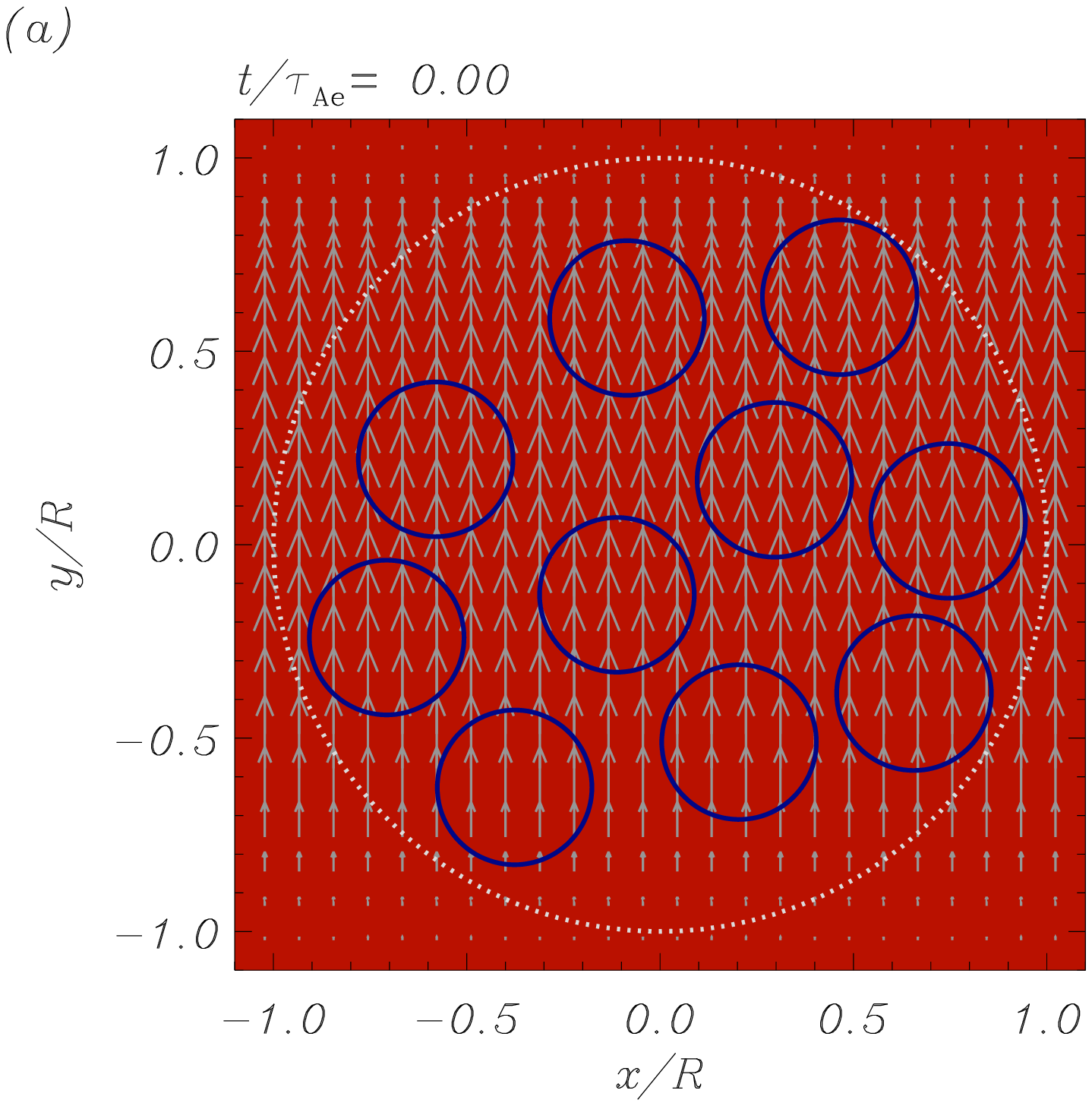}\hspace{-0.cm}\includegraphics[
width=7.cm]{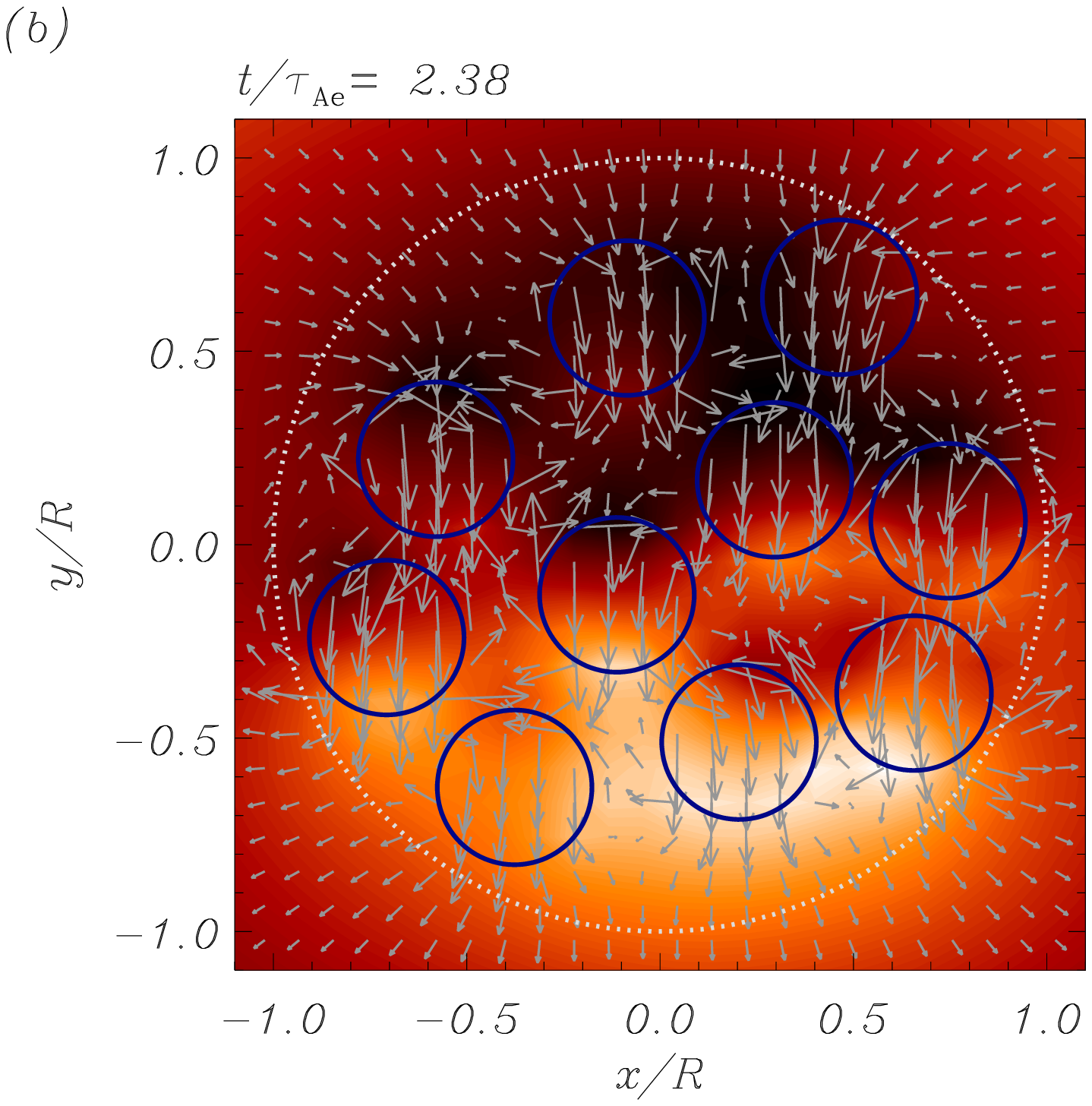}\vspace{-0.5cm}
\includegraphics[width=7.cm]{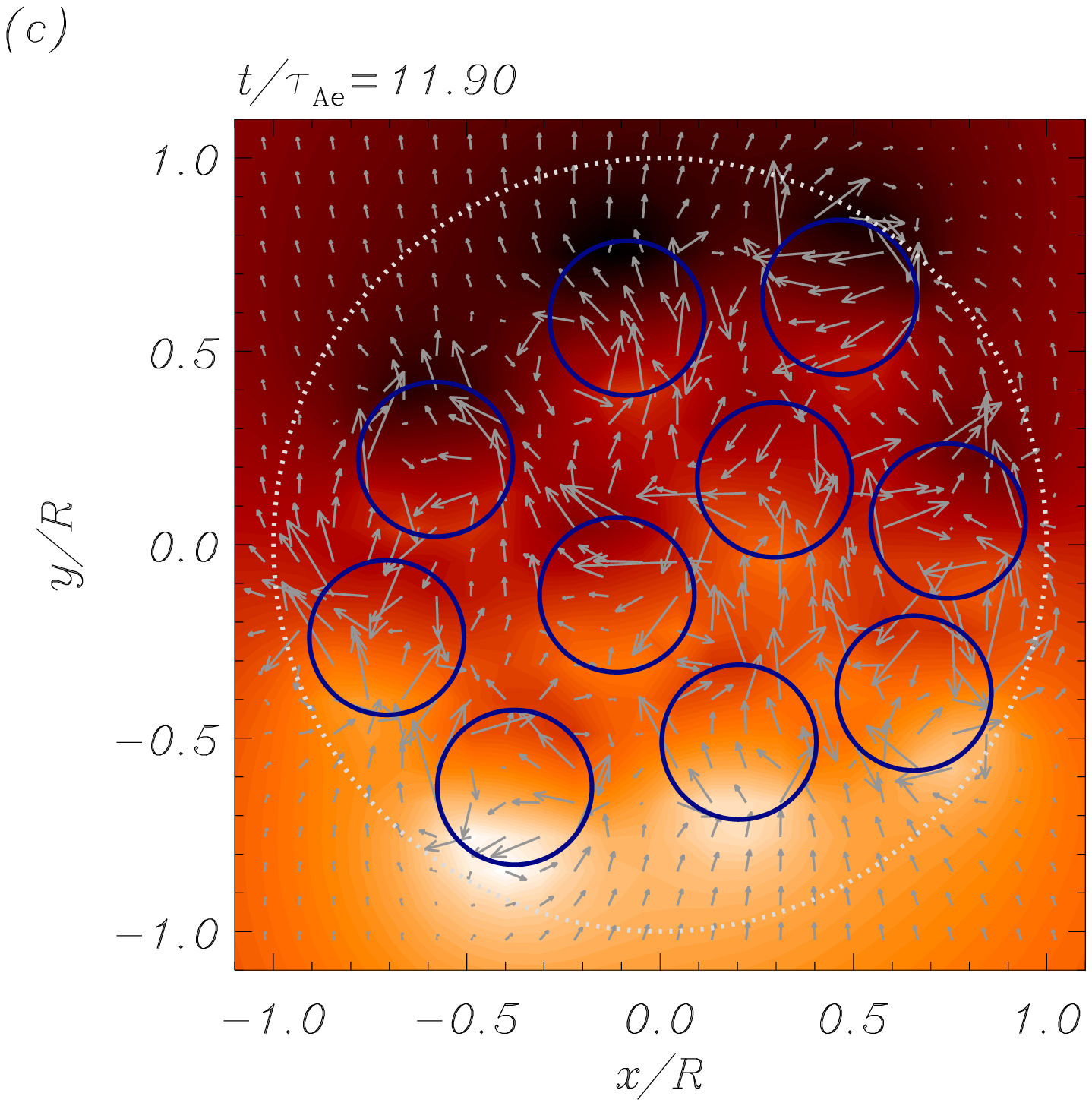}\hspace{-0.cm}\includegraphics[
width=7.cm]{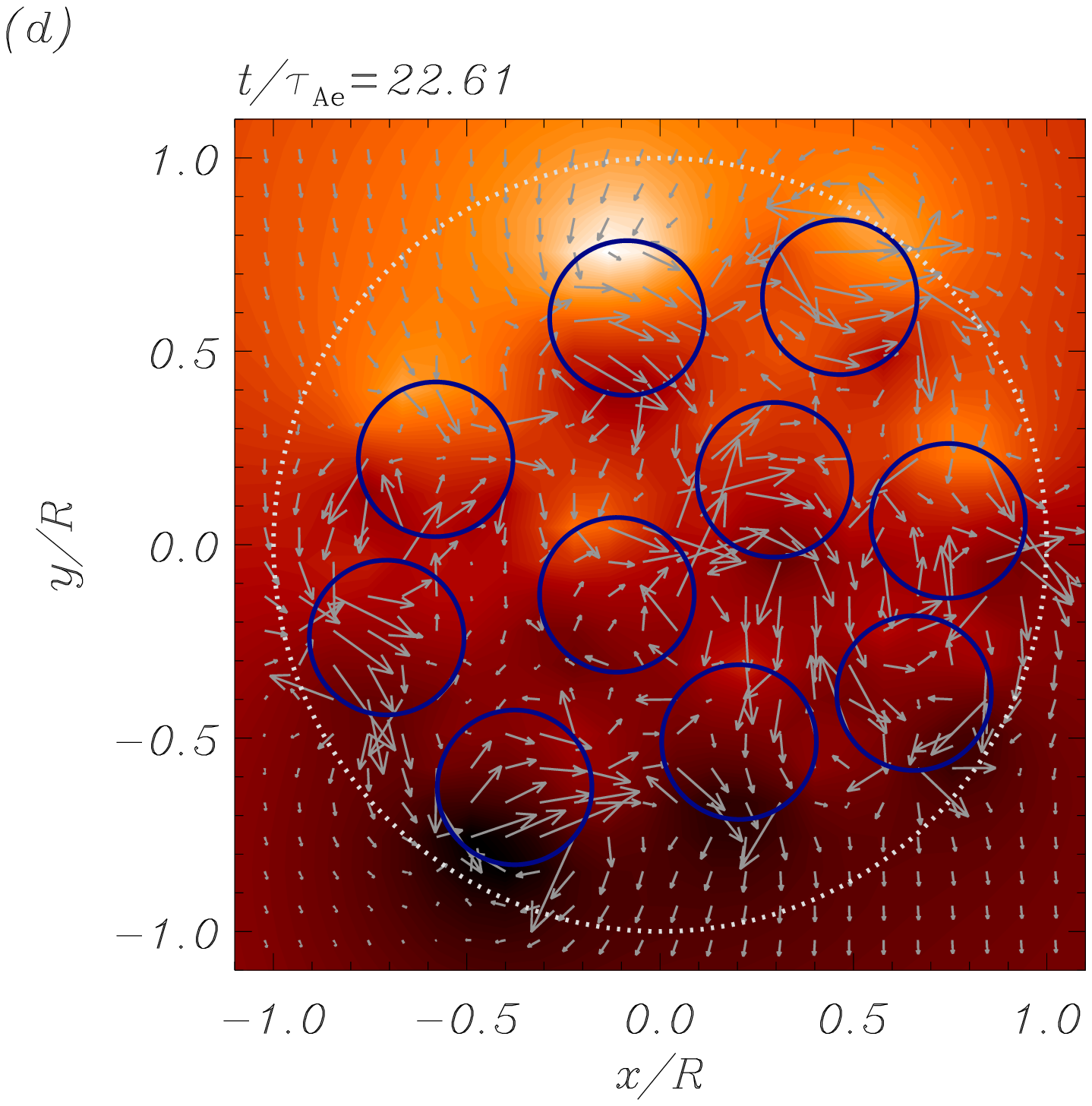}\vspace{-3.cm}
\includegraphics[width=7.cm]{bar.eps}\vspace{-2.5cm}
\caption{Time-evolution of the velocity field (arrows) and magnetic pressure
perturbation (colored contours) for the system of $10$ identical strands of \S
\ref{10id_strands}. The panels show different evolution times. In {\bf (a)} the
initial condition (Eq. \ref{initial_condition}) of the velocity field is
plotted. In {\bf (b)} the magnetic pressure and velocity fields are shown
shortly after the initial disturbance. The initial pulse has left the domain
shown in this panel and the system oscillates coherently. In {\bf (c)} and {\bf
(d)} the structure of the fields indicate a complex motion of the strands. In
{\bf (c)} the direction of oscillation is completely different from that of the
initial disturbance and in {\bf (d)} the transverse displacement of the strands
is mainly in the direction perpendicular to that of the initial pulse.}
\label{simulation01}
\end{figure}

The initial perturbation is a planar pulse in the velocity field of the form
\begin{equation}\label{initial_condition}
\mathbf{v}_{0} = V_{0}~e^{-\left(y / w_0\right)^2} \mathbf{e}_y,
\end{equation}
(see arrows in Figure \ref{simulation01}(a)) where $w_0$ is the width of the
Gaussian profile, and $V_{0}$ is its amplitude. We have set the width of
the initial perturbation equal to the
loop radius, $w_0=R$, to perturb all the strands (see Figure
\ref{simulation01}(a)). In addition, we have chosen an amplitude of the
perturbation, $V_{0}=0.02 \,v_{\mathrm{A e}}$, such that the maximum
displacement of each strand is equal to the radius, $a=0.2 R$. The initial
value of the $x$-component of the velocity
and the magnetic field perturbation are zero. Thus, the magnetic pressure is
initially zero. We numerically solve the initial value problem with a code
developed by J. Terradas based on the Osher-Chakrabarthy family of linear flux
modification schemes \citep[see][]{bona2009}. The size of the simulated domain
is $2R\times2R$ and its boundaries are sufficiently far to neglect the effects
of the reflections on the multi-stranded loop dynamics. The numerical mesh has
$1000\times1000$ grid-points and has enough resolution to resolve small scales
and to avoid significant numerical diffusion. 

Figure \ref{simulation01} shows the temporal evolution of the magnetic pressure
and velocity fields. The initial disturbance excites the $v_{y}$ component and
the pulse front is on the $x$-axis (see Figure \ref{simulation01}(a)). All the
strands are excited at the same time and this produces the motion of the whole
loop in the positive $y$-direction. In Figure \ref{simulation01}(b), part of the
perturbation energy has leaked from the system during the transient phase. The
strands oscillate in the negative $y$-direction and in phase, in some kind of
global-kink motion. The velocity field has a relatively simple structure, having
a uniform value inside the strands. After some time, the spatial structure of
the velocity and magnetic pressure perturbation fields are more complex (see
Figure \ref{simulation01}(c)). The polarization of the strand motion no longer
parallel to the $y$-axis and each strand oscillates in its own direction.
Similarly, in Figure \ref{simulation01}(d) the velocity and magnetic pressure
fields have also a complicated structure and the direction of oscillation of
each strand has changed from that of Figure \ref{simulation01}(c). Then, the
initial value problem shows that the complexity of the magnetic pressure and
velocity fields increases in time and that the simple spatial structure of
Figure \ref{simulation01}(b) is not recovered after the initial stage.

\begin{figure}[!ht]
\centering
\includegraphics[width=12.cm]{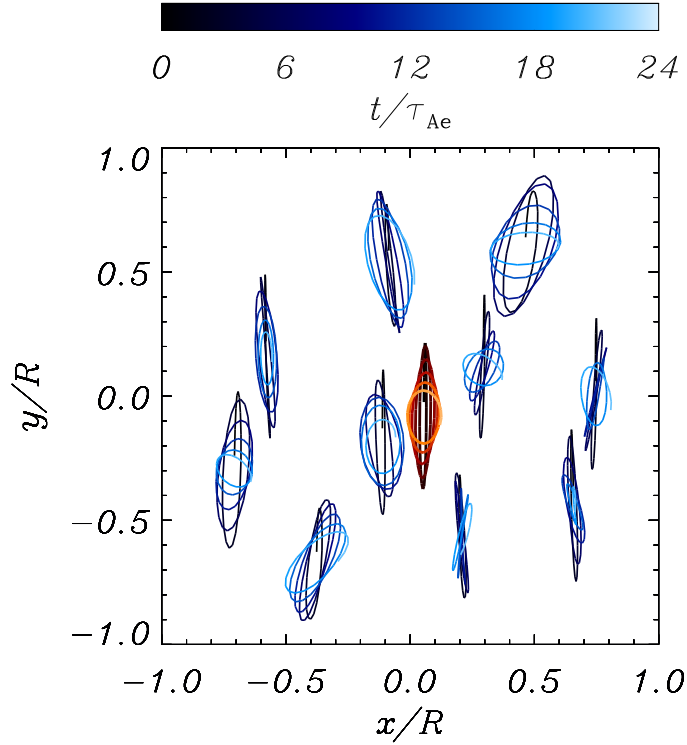}
\caption{Plot of the trajectories of the strand centers (blue solid line) and
the trajectory of the center of mass of the system (red solid lines). Time is
represented by the color lightness of the curves start with dark color and end
with light color (see color bar). The color bar associated to the motion of the
center of mass is not plotted for the sake of simplicity. The strands and the
center of mass start moving roughly parallel to the $y$-axis, but after some
time the trajectories are ellipsis. To clarify the movements of the
strands,
the displacements are multiplied by two in this plot.}
\label{trajectories}
\end{figure}

\begin{figure}[!ht]
\centering
\includegraphics[width=8.cm]{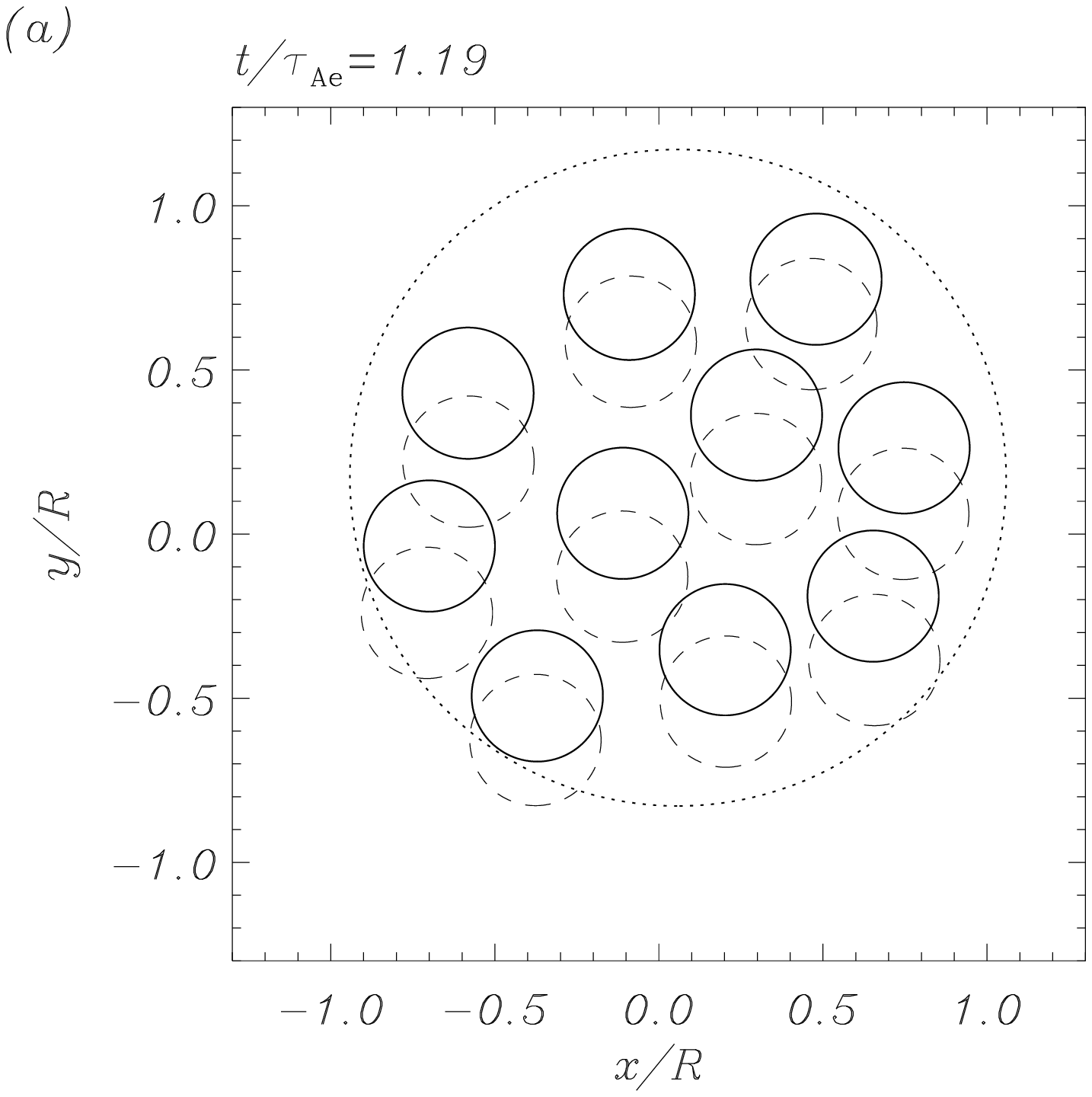}\includegraphics[width=8.cm]{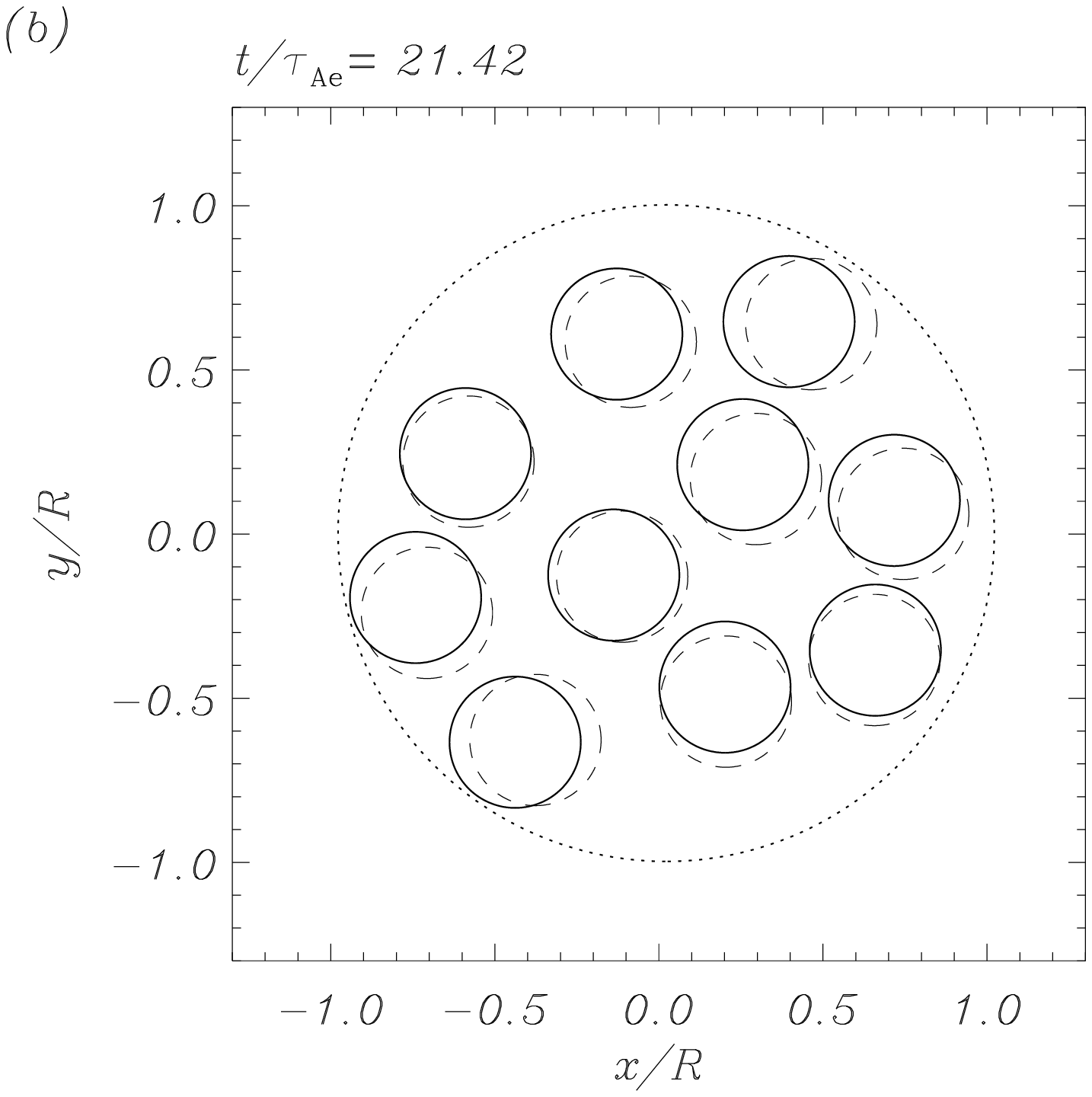}
\caption{Time-evolution of the displacement of the strands (solid circles)
and monolithic loop (large dotted circle). The initial position of the strands
is
also plotted as dashed circles. In {\bf (a)} the maximum displacements of the
strands exerted by the initial condition are plotted. All displacements are in
the positive $y$-direction and in phase, indicating a coherent motion of the
strand set. In {\bf (b)} there is no privileged direction of oscillation, and
complex motions of the strands are shown. The time of the two snapshots is
displayed at the top of the figures.
This figure is available as an mpeg animation in the electronic edition of
\emph{The Astrophysical
Journal}. }
\label{displacements}
\end{figure}

The temporal evolution of the velocity field indicates a complex motion of the
strands. To show this more clearly, the trajectories of the strand centers are
plotted with colored blue lines in Figure \ref{trajectories}. Initially, all the
strands oscillate in the $y$-direction, i.e., the direction of the initial
disturbance. After a short time, the direction of the transverse oscillation of
each strand changes and complicated trajectories arise. The motion of each
strand produces a strong modulation of the whole loop transverse displacement.
The trajectory of the center of mass (defined as
$\mathbf{r}_\mathrm{CM}=\sum_{j=1}^{N} \mathbf{r}_{j}/N$) is plotted in Figure
\ref{trajectories} as a colored red line. This trajectory represents the whole
loop motion and it shows two effects. The first effect is that the initial
linear polarization of the loop oscillation changes to a circular polarization
in which the loop orbits around a central position. The second effect is an
attenuation of the oscillation. The reason of this attenuation is that the
non-organized motions contribute less to the whole loop motion than the initial
organized motions. This behavior is even clearer in Figure
\ref{displacements} and the movie available in the electronic edition of the
Journal. This figure shows the
temporal evolution of the displacements of the strands with respect to their
initial positions and also the displacement of the hypothetical monolithic
loop.

\begin{figure}[!ht]
\centering
\includegraphics[width=10cm]{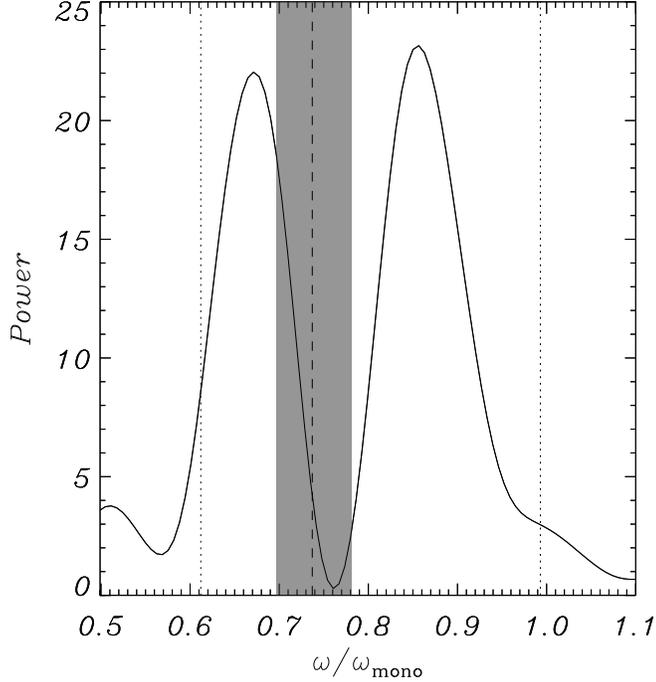}
\caption{Power spectrum of the temporal evolution of the magnetic pressure
perturbation measured at the position $(x/R,y/R)=(0.15,0.31)$. The left and
right vertical dotted lines represent the frequencies of the lowest and highest
frequency modes of Figure \ref{eigenfrequencies}(a), respectively. The vertical
dashed line is the kink frequency of the individual strands,
$\omega_\mathrm{strand}$, while the shaded area, that corresponds to the shaded
area
in Figure \ref{eigenfrequencies}(a), marks the region where mid modes reside.
The power spectrum has two peaks in the low and high mode frequency ranges,
which implies that the initial disturbance mainly excites the low and high
frequency modes.}
\label{periodogram_gap}
\end{figure}

In Figure \ref{periodogram_gap}, we have plotted the power spectrum of the
magnetic pressure perturbation measured in a point located in the fluid between
the strands. This figure shows that all the power is concentrated in the
frequency band of the collective normal modes (see Figure
\ref{eigenfrequencies}(a)). This indicates that the initial disturbance excites
a combination of collective normal modes (see \S \ref{10id_strands}). In
general, the particular combination of normal modes depends on the shape,
position, and incidence angle of the initial pulse \citep[see][]{luna2008}. With
the particular initial disturbance of Equation (\ref{initial_condition}), the
power spectrum has the form of two peaks centered in the low and high frequency
regions, while the power in the mid frequency region is small. Then, the initial
disturbance largely excites the low and high frequency modes.

\citet{ofman2008} reported transverse oscillations of a multi-stranded
loop. The system is made up of several close strands, and one may expect that
the strands interact and oscillate with a combination of collective normal
modes. Assuming that the system of strands is similar to that of \S \ref{10id_strands}, we can estimate the range of periods of collective normal modes. We use the mean values of the density and magnetic field obtained by the authors of $3\times10^{9}~\mathrm{cm^{-3}}$ and $20~\mathrm{G}$, respectively. Also, imposing a reasonable external density of $4\times10^{8}~\mathrm{cm^{-3}}$ \citep[see][]{aschwanden2003}, we keep the density contrast of our model ($\rho_{j}/\rho_\mathrm{e} = 7.5$). With these values, the estimated collective periods range from $94~ \mathrm{s}$ to $152~ \mathrm{s}$, in good agreement with the $113~\mathrm{s}$ of the averaged oscillating periods of the strands measured by the authors. On the contrary, if the same system is assumed as a monolithic loop oscillating with a period of $113~\mathrm{s}$, the estimated magnetic field is $15~\mathrm{G}$. This very preliminary result shows that observations with poor spatial resolution tend to underestimate the magnetic field.

\section{Discussion and conclusions}\label{dis_conc}

In this work we have studied analytically the normal modes of a multi-stranded
coronal loop in the $\beta=0$ limit with the help of the $T$-matrix theory. We
have also studied the temporal evolution of the system after an initial
disturbance and its relation with the normal modes. The results of this work can
be summarized as follows

\begin{enumerate}
\item We have considered a multi-stranded loop filled with $10$ identical
strands located at random positions. We have found that the system supports a
large quantity of normal modes whose frequencies are in a broad band of width
approximately $0.38\omega_\mathrm{mono}$. All these frequencies are smaller than
the monolithic kink frequency. The collective normal modes can be classified in
three groups according to their frequencies and spatial structures. Low modes
have a frequency $\omega\lesssim\omega_\mathrm{strand}$ and the spatial
structure is kink-like and characterized by strands moving in complex chains. In
these modes the intermediate fluid between strands follows their transverse
displacement and this produces a non-forced motion of the system. In the low
modes the strands move faster than the surrounding medium, i.e., the maximum
velocities are within the strands. Mid modes have a frequency
$\omega\approx\omega_\mathrm{strand}$ and the spatial structure is fluting-like,
by which the strands are essentially distorted and their transverse
displacements are small. Finally, high modes
($\omega\gtrsim\omega_\mathrm{strand}$) are kink-like modes characterized by a
forced motion of the strands, that move in the opposite direction to the
surrounding plasma or compress and rarefy their intermediate fluid, producing
high velocities in the coronal medium. Then, the surrounding medium moves faster
than the strands. 

\item We have also investigated a system of $10$ non-identical strands. The
spatial distribution of the strands is the same as in \S \ref{10id_strands}, but
with different strand densities. Similarly to the identical strand case, we have
found a large quantity of collective normal modes, but now their frequencies lie
in a band of width $0.30\omega_\mathrm{mono}$. This band width is narrower than
that of the identical strand case of \S \ref{10id_strands}, indicating a weaker
interaction between the strands. The collective normal modes can be also
classified in low, mid, and high modes. The largest oscillation amplitudes
correspond to the denser strands in the low modes and to the rarest strands in
the high modes.

\item The normal modes of a complex system of $40$ identical strands have also
been computed. Their frequencies lie in a band of width
$0.37\omega_\mathrm{mono}$ that coincides well with that of the system of $10$
identical strands. The classification of the normal modes in low, mid, and high
is still valid in this complex system, although the number of normal modes is
larger than in the two systems with $10$ strands. This indicates that the number
of collective normal modes increases with the number of strands. In addition,
these results indicate that the width of the frequency band does not depend
strongly on the number of strands.

\item The temporal evolution of the system after an initial planar disturbance
is also studied in the system of $10$ identical strands. Initially, the system
oscillates in phase in the direction of the initial disturbance. After some
time, this organized motion disappears and the complexity of the velocity and
magnetic pressure perturbation fields increase. This implies a complex motion of
the strands and, as a result, of the whole loop. In addition, we have found that
the system oscillates with a combination of low and high collective normal
modes.

\end{enumerate}

In this investigation, we show that the transverse oscillation of a
multi-stranded loop
cannot be described by an equivalent monolithic loop. The reasons are that there
is a huge quantity of normal modes with very different frequencies and very
complex spatial structures. Their frequencies lie in a broad band and cannot be
accounted for by an average frequency, because, after an initial disturbance
most of the frequencies are excited. Furthermore, there is no collective normal
mode that can be considered as a global kink mode, in which all the strands move
in phase with the same direction and produce a transverse displacement of the
whole loop. Instead, the collective normal modes that we have found displace the
loop center but the detailed motion of the strands is very complex.

Additionally, the internal fine structure influences the whole loop dynamics.
Complex motions of the strands are produced and also complex movements of the
whole loop. These motions can be explained by the existence of a strong
interaction between the strands and, consequently, the existence of a huge
quantity of collective normal modes with different frequencies. The initial
disturbance excites a particular combination of modes that causes a coherent
motion of the strands similar to a global kink transverse oscillation. After
some time, each collective normal mode oscillates with different phase due to
the frequency differences between them. As a consequence the coherent motion of
the strands is lost and complex motions appear. This behavior has already
been shown by \citet{luna2008} in a system of two identical flux tubes. The
change of the initial linear polarization to a circular polarization of the
whole loop transverse oscillation may be a
signature of its internal fine structure. Circular transverse
motion of a loop has been reported by \citet{aschwanden2009}, who reconstructed
the 3D motion by the curvature radius maximization method from TRACE images
taken in the loop oscillation event of 1998 July 14. The author found that the
horizontal and vertical oscillations have similar period and a phase delay of a
quarter of a period. We suggest that the internal (thread) structure can
contribute to the circular polarization and to the rapid damping of the
transverse oscillations of coronal loops. We also show that the magnetic
field strength tends to be underestimated in an observation of a multi-stranded
loop oscillation in which the internal fine structure is not resolved. This
result supports the findings of \citet{demoortel2009}, who found that the
estimated local magnetic field strength strongly depends on the theoretical
model used to compare with the observations. Better models of coronal loop
oscillations will improve the accuracy and reliability of the estimated
magnetic fields obtained with the coronal seismology method. However,
this is a preliminary study and more observations of circular transverse
motions, a detailed study of the relation between the damping and the internal
fine structure, and more high resolution measurements are needed. The
recently launched Solar Dynamics Observatory (SDO) will provide new data with
high spatial and temporal resolution. With these new observations better models
of multi-stranded loops will be done.

In this work, we have made a number of simplifying assumptions, neglecting
gas pressure, considering only linear perturbations, and ignoring gravity. In
order to have more realistic models these effects need to be incorporated.
\citet{soler2009} have studied a system of two prominence threads with gas
pressure and have found that transverse oscillations are coupled. These authors
have also shown that slow modes are essentially individual modes. Then, we
expect that the results shown here are still valid in a system with finite beta.
Nevertheless, the addition of non-linear terms and gravity may introduce new
effects that need to be addressed in a future research.

\acknowledgments 

M.L. is grateful to the Spanish Ministry of Science and Education for an FPI
fellowship, which is partially supported by the European Social Fund. The
authors acknowledge the Spanish Ministry of  Science and Education and the
Conselleria d'Economia, Hisenda i Innovaci\'o of the Goverment of the Balearic
Islands for the funding provided under grants AYA2006-07637, PRIB-2004-10145,
and PCTIB-2005-GC3-03, respectively. M.L. also thanks the partial support of
NASA under contract NNG06EO90A. The authors thank an anonymous referee for useful suggestions that helped improve this paper.

\end{document}